\documentclass[prl,onecolumn,superscriptaddress,amsmath,amssymb]{revtex4} 
\usepackage{appendix}
\usepackage{graphicx}
\usepackage{epsfig} 
\usepackage{caption}
\usepackage{subfloat}
\usepackage[justification=raggedright]{caption}
\usepackage[dvipsnames]{xcolor}
\usepackage[T1]{fontenc}
\usepackage{wrapfig}
\usepackage[english]{babel}
\usepackage{mathrsfs} 
\usepackage{amsmath}
\usepackage{amsfonts}
\usepackage{sidecap}
\usepackage{bbm}
\usepackage[multidot]{grffile}
\usepackage{subfig}
\usepackage{multirow}
\usepackage{url}
\usepackage[english]{babel}
\usepackage{algorithm}
\usepackage{algorithmic}
\usepackage{amsmath,bm}
\newcommand{\carla}[1]{\textcolor{black}{#1}}
\newcommand{\red}[1]{\textcolor{black}{#1}}

\def\a{\alpha}

\def\d{\delta}

\def\s{\sigma}

\def\dd{\dagger}

\def\mean#1{\left< #1 \right>}
\DeclareMathOperator{\Tr}{Tr}

\begin{document}
\title{Maximally Localized Dynamical Quantum Embedding for Solving Many-Body Correlated Systems}
\author{Carla Lupo}
\affiliation{King's College London, Theory and Simulation of Condensed Matter, The Strand, WC2R 2LS London, UK}
\author{Fran\c{c}ois Jamet}
\affiliation{King's College London, Theory and Simulation of Condensed Matter, The Strand, WC2R 2LS London, UK}
\affiliation{National Physical Laboratory, Teddington, TW11 0LW, United Kingdom}
\author{Terence Tse}
\affiliation{King's College London, Theory and Simulation of Condensed Matter, The Strand, WC2R 2LS London, UK}
\author{Ivan Rungger}
\affiliation{National Physical Laboratory, Teddington, TW11 0LW, United Kingdom}
\author{Cedric Weber$^*$ }
\affiliation{King's College London, Theory and Simulation of Condensed Matter, The Strand, WC2R 2LS London, UK}

\begin{abstract}
We present a quantum embedding methodology to resolve the Anderson impurity model in the context of dynamical mean-field theory, based on an extended exact diagonalization method. Our method provides a maximally localized quantum impurity model, where the non-local components of the correlation potential remain minimal. This method comes at a large benefit, as the environment used in the quantum embedding approach is described by propagating correlated electrons and hence offers a \red{polynomial increase $O(N^4)$ of the number of degrees of freedom for the embedding mapping without adding bath sites}.  
We report that quantum impurity models with as few as 3 bath sites can reproduce both the Mott transition and the Kondo physics, thus opening a more accessible route to the description of time-dependent phenomena. Finally, we obtain excellent agreement for dynamical magnetic susceptibilities, poising this approach as a candidate to describe 2-particle excitations such as excitons in correlated systems. We expect that our approach will be highly beneficial for the implementation of embedding algorithms on quantum computers, as it allows for a fine description of the correlation in materials with a reduced number of required qubits.
\end{abstract}

\maketitle

\thispagestyle{empty}


\section*{Introduction} 


The understanding of materials with strongly correlated electrons is one of the main challenges of modern solid state physics. Triggered by the discovery of high-temperature superconductivity in copper-oxides, the study of doped Mott insulators has grown in the last decades, building on the development of theoretical tools designed to solve models of strongly correlated electrons accurately. 
Despite lacking in terms of the exact solution of simple correlated theoretical models in two or three dimensions, accurate predictions for the properties of strongly correlated solids are obtained by using approximations \cite{kotliarRMP}. A central role has been played by the dynamical mean field theory (DMFT)\cite{rmp}, a non-perturbative method that allowed for the first complete description of the Mott-Hubbard transition. This method has been extended to a variety of correlated methods and combined with density functional theory, 
leading to remarkable agreement with the properties of many correlated materials. DMFT
more generally falls within the larger group of quantum embedding theories 
\cite{embedding_seet, dmet_edoardo} (for a review, see Ref.[\cite{chan_embedding}] and references therein), which have been widely successful at describing transition metal and $f$-elements into both solids and molecular systems. 
The central idea of DMFT is to self-consistently map the infinite bulk system onto a so-called Anderson impurity model (AIM) with only a few interacting impurity sites embedded in an infinite non-interacting bath. The latter Anderson impurity models can be solved using high-level many-body methods, with a breadth of approaches for all have their own limitations.
Indeed, the recent development of CTQMC has generated strong activity in the field. For single-site DMFT, CTQMC yields an exact solution to the AIM problem within the statistical error bars in imaginary time.  The main limitation of the approach is that the evaluation of real-frequency spectra requires a poorly conditioned analytical continuation based on the maximum entropy method\cite{mem,Gunnarsson10} (or some alternative strategy), which strongly limits the possibility to study fine details of the spectra. 
For multi-orbital or cluster extensions of DMFT, CTQMC suffers from the \emph{fermionic sign-problem} 
as long as inter-orbital hybridizations are present, and it is therefore limited to finite temperatures.

Numerical renormalization group (NRG) \cite{andrew_kondo_nrg} provides an alternative for real axis calculations and access to Kondo physics, but remains challenging to extend for multi-orbital systems. 

\begin{figure}[]
\begin{center}
\includegraphics[width=0.7\columnwidth]{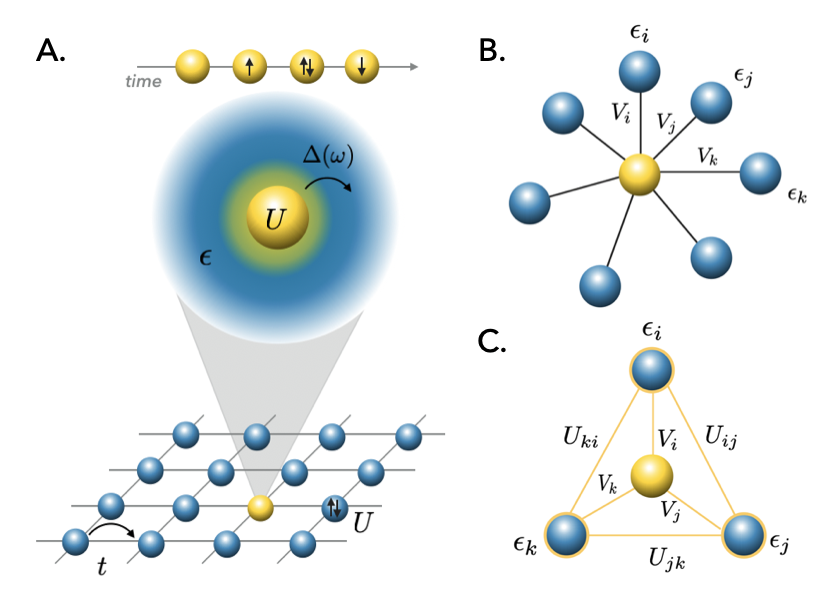}
\caption{\textbf{From AIM to MLDE}. (a) Mapping of the lattice model (with inter-site hopping parameter t and on-site correlation U) onto a local impurity model, with a correlated atom (yellow sphere) embedded in a non-interacting bath (blue shaded area) with energy $\epsilon$ as for the Anderson impurity model (AIM). The four possible configurations describe the quantum evolution of the atom. Electrons may hop from the atom to the bath via the frequency dependent hybridization function $\Delta(\omega)$, which plays the role of a dynamical mean field. b) Discretization of the continuum bath in non-interacting bath sites coupled to the interacting impurity through the hopping function $V_i$. This is the picture related to the Exact Diagonalization (ED) impurity solver. c) Cartoon related to the Maximally Localised Dynamical Embedding (MLDE) solver where an interacting impurity is coupled to interacting bath sites.} \label{fig:benchmark_2}
\end{center}
\end{figure}

Finally, exact diagonalization (ED) solvers are instead based on a finite discretization of the AIM through representation of the effective bath in terms of a small number of \emph{bath-sites}.
In practical implementations, the bath size ($N_b$) is severely limited because of the growth of the Hilbert space: its dimension scales exponentially with the total number of sites $N_s$ (bath sites and impurity orbitals). Nonetheless, the use of Lanczos-based algorithms 
allows to deal with large Hilbert spaces, and the discretization at low temperature \cite{lucaED,leibsch_ed_lanczos_dmft} is fine enough to compute the thermodynamic and static observables accurately.
In particular, the finite size effects affect the spectral functions, which are slowly converging to the continuous features of the exact DMFT solution.
Notwithstanding reasonably accurate static phase diagrams obtained with the Lanczos solver for three \cite{Capone02,ed_solver_liebsch_review} and five \cite{Ishida10} orbitals system, the limitation in the bath size also becomes particularly relevant for multi-orbital AIM models, which are necessary to realistically describe transition metal oxides, as the impurity's orbitals (five for a $d$-manifold) contribute to the enlargement of the Hilbert space. \red{Toward the achievement of the largest number of bath sites, ED calculations have been extended to handle the single impurity embedding problem, allowing from O(100) \cite{Arrigoni_2013,Lu_DMFT_2014,Ganahl_2015,Bauernfeind_2017} up to O(300) \cite{Lu_2014} uncorrelated bath sites.}
\red{Although these approaches allow the obtaining of approximations of the zero temperature Green’s function, the building of systematically high energy excited states remains challenging.}
Moreover, scaling the precision of the latter approach on classical computers is not feasible due to the exponential increase of the Hilbert space with the number of bath sites. In view of the recent progress in the implementation of ED on quantum computers \cite{ivan_quantum_computing}, the latter approach has gained interest due to the possibility of scaling linearly on quantum computers. However, due to current hardware limitations (noise and decoherence issues), such algorithm are limited to a small number of qubits and short quantum circuits on the currently available noisy intermediate-scale quantum (NISQ) computers. Hence it is of great importance to develop an ED based solver that can obtain precise results with a reduced number of bath sites, since the required number of qubits is proportional to this.

In this light, we present here a new quantum embedding methodology to resolve the Anderson impurity model (AIM) for DMFT, based on an extended ED method. Our method provides a maximally localized quantum impurity model, where the non-local self-energy component of the correlation remains minimal, and hence the AIM minimally breaks locality (DMFT is a purely local theory). As reported in this work, this comes at a remarkable benefit, as the environment used in the quantum embedding approach is described by propagating correlated electrons (instead of free electrons in DMFT) \red{and hence offers an polynomial increasing of the numbers of degrees of freedom for the embedding mapping, without increasing the number of bath sites.} 
This is reminiscent of the representation of correlated electrons by a Green's function embedding approach, where correlations are described by hidden fictitious additional fermionic degrees of freedom \cite{hidden_fermions,andrew_topology_mapping}. 

This representation has hence the potential to improve the scope of applicability of the quantum embedding approach dramatically, whilst limiting the small number of bath sites. 
We report that quantum impurity models with as few as 3 bath sites can reproduce both the Kondo regime and the Mott transition and obtain excellent agreement for dynamical magnetic susceptibilities, poising this approach as a candidate to describe 2-particle excitations such as excitons in correlated systems, such as high-Tc superconductors \cite{apical_cedric_phonon,apical_cedric_prx}. Our approach aligns with recent progress in quantum computing, where a realistic number of qubits would achieve a fine description of correlations in materials.

\section*{Results}

\subsection{From the AIM to MLDE}
Within DMFT, the  lattice model is mapped to an effective Anderson impurity model (AIM)  where a correlated atom is connected to a non-interacting bath with the hybridization function $\Delta(\omega)$ as shown in Fig.1.a.
In the exact diagonalization (ED) approximation, the continuum bath is represented with a finite number of effective sites (Fig.1.b).
Typically, for a fixed set of the Hamiltonian parameters (see Eq.1 Methods section), the AIM is solved (\ref{aim}) by using a Lanczos algorithm to converge the ground state and excited states  \cite{lucaED,ed_solver_liebsch_review} which contribute to the thermal average. Once the eigenstates are obtained, the dynamical and static observables are computed.
As previously stated, the number of bath sites is severely limited because of the exponential scaling of the Hilbert space with the number of sites. 
To improve the performance of the algorithm, we enlarge the degrees of freedom of
the approach introducing a two-body interaction between the bath electrons (Fig.1.c). The resulting approximation is represented by an extended ED solver where the non-local component of the correlation potential remain minimal. Thus the exact diagonalization method has been extended to a maximally localized dynamical embedding (MLDE) model where the same accuracy is obtained with a reduced number of bath sites. (see Methods section for the formal mathematical definition of the model).

\begin{figure}[!h]
\begin{center}
\includegraphics[width=0.7\columnwidth]{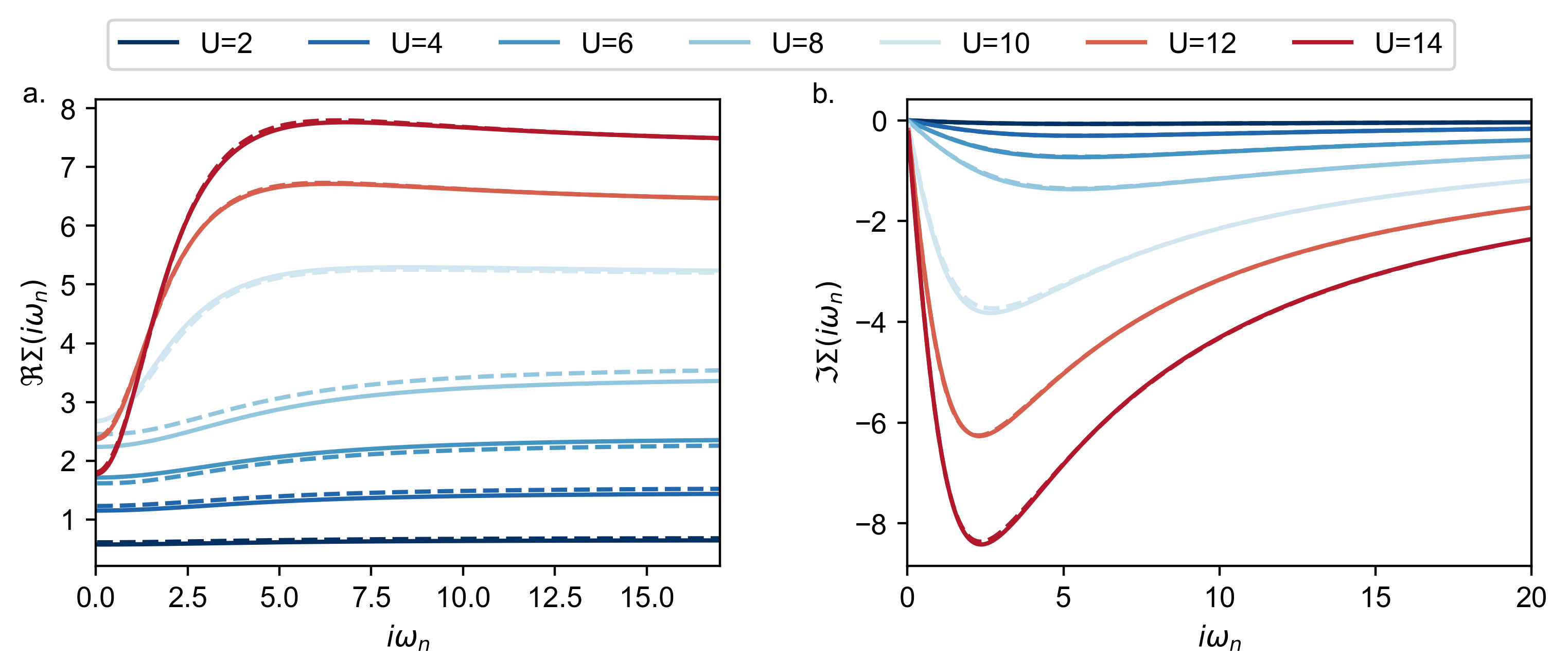}
\caption{\textbf{MLDE benchmark}. (a) Imaginary and (b) real part of the MLDE with $N_b=3$ self-energy (dashed lines) obtained from on-site correlation $U$ in range $2-15$ for a test hybridisation function for a half-filled impurity (units are arbitrary) and compared with continuous
time Monte Carlo (continuous line). The scales are in arbitrary units.} \label{fig:aim}
\end{center}
\end{figure}

\begin{figure}[!h]
\begin{center}
\includegraphics[width=0.7\columnwidth]{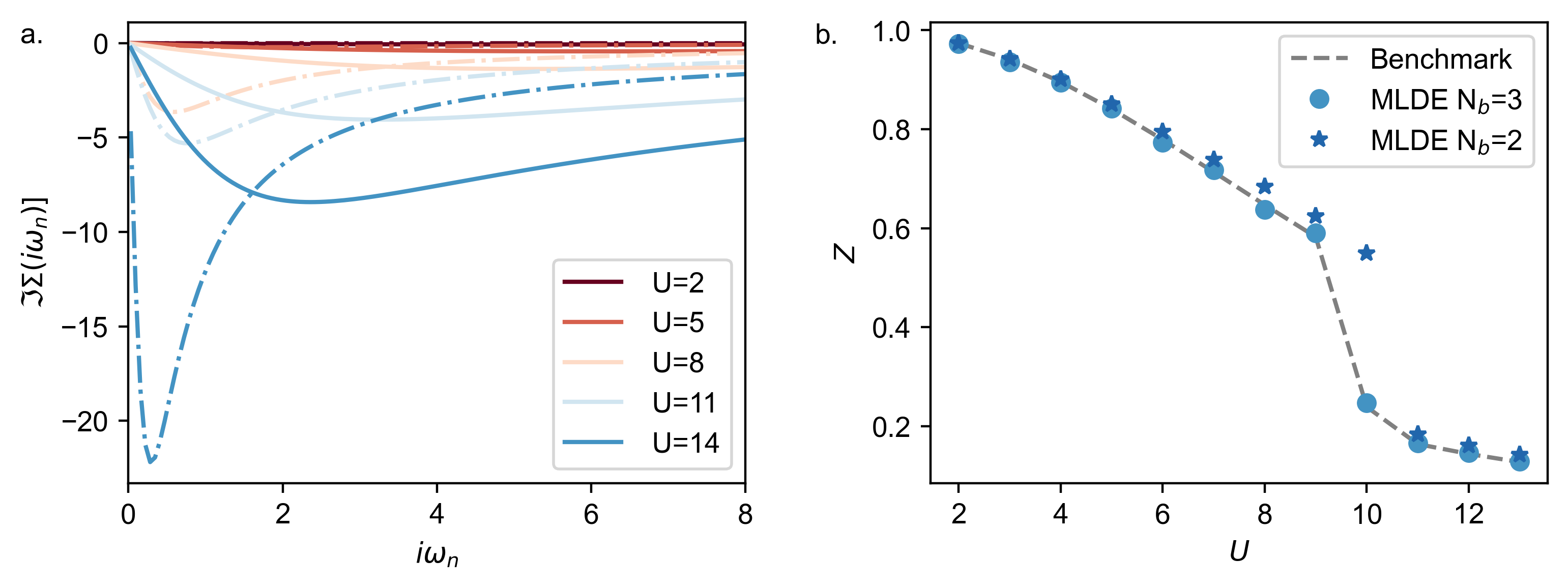}
\caption{\textbf{Strength of correlations}. (a) Imaginary part of the self-energy 
obtained by ED for $N_b=2$ (dashed lines) compared to the exact solution (continuous line). (b) Quasi-particle weight obtained by MLDE with $N_b=2$ and $N_b=3$ compared to the continuous time Monte Carlo. The MLDE remains
close to the exact solution with $N_b=2$, whereas for the same number of bath sites ED vastly overestimates the strength of correlations (see panel (a)). 
The scales are in arbitrary units.} \label{fig:benchmark}
\end{center}
\end{figure}

\subsection*{MLDE and CTQMC}
\label{sec:mlde_ctqmc}
We now turn to a simple test of the MLDE equations for a typical AIM. The bath hybridization used as a test case is obtained from a typical correlated material \footnote{The hybridization function is available in the supplementary information, Fig.S1}, and the impurity energy is set to $\epsilon_f=-U/2$ to stay at half-filling.
We perform a benchmark of the MLDE approach with respect to both a continuous-time Monte Carlo and Lanczos solver, which both provide in this case the same answer used as a reference. Both the imaginary part (Fig.\ref{fig:aim}.a) and real part ( Fig.\ref{fig:aim}.b) obtained by MLDE with as few as $N_b=3$ bath sites provide a remarkable agreement with the exact solution. We considered both the nearly free electron (NFE) limit ($U<10$) and the atomic limit ($U>10$), and in both cases the MLDE solution is consistently in agreement with the exact answer. Noteworthy, the strength of correlation and overall physical properties are also well captured by MLDE with $N_b=2$, whereas the ED solver with the same number of bath sites largely overestimates the strength of correlation (see Fig.\ref{fig:benchmark}.a). MLDE captures well the delocalization-localization transition (see Fig.\ref{fig:benchmark}.b), where the MLDE with $N_b=2$ only differs near the transition while MLDE with $N_b=3$ is exact.

\begin{figure}[!t]
\begin{center}
\hspace*{-0.4in}
\includegraphics[width=0.7\columnwidth]{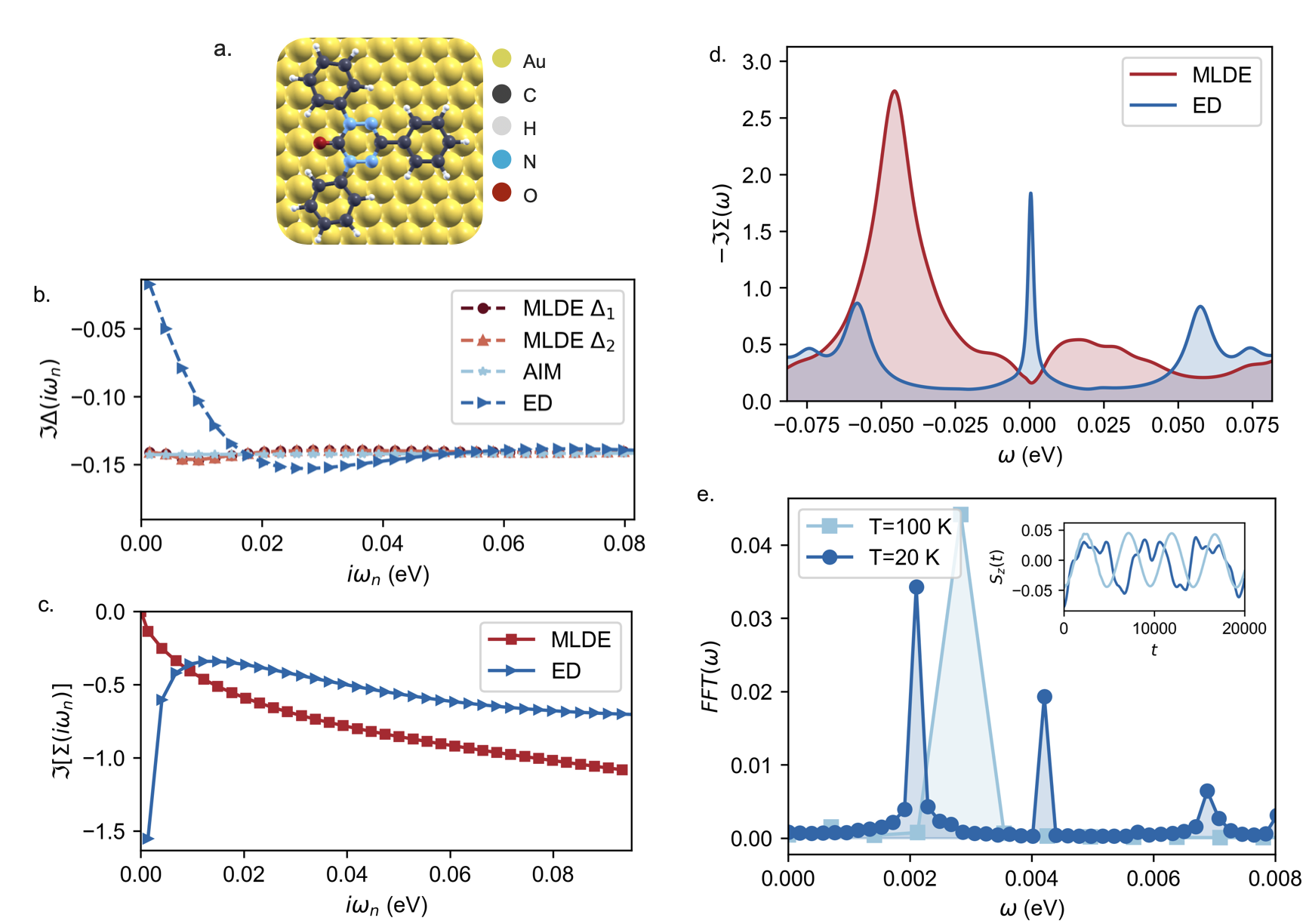}
\caption{\textbf{Kondo physics}. (a) TOV organic molecule deposited on a gold substrate at $T=5$K. (b) Imaginary parts of the molecule's hybridisation function (AIM), represented by MLDE $\Delta_{1,2}(i\omega_n)$, and as obtained by the ED solver with $N_b=12$. In the Kondo regime, the hybridisation remains constant up to the lowest frequency, which is not captured in ED (c) Imaginary part of the self-energy obtained by MLDE shows a Fermi liquid type behaviour at small frequency (squares), whereas the self-energy shows an artificial Mott singularity, due to the bath
discretization (truangles). We note that this low temperature was beyond the reach of our CTQMC solver ($\beta=35000$). (d) Imaginary part of the self-energy in real axis frequencies, MLDE shows a small
dip near the Fermi level, which is associated with the Kondo scale, whereas ED shows an artificial peak instead. (e) Time dynamics of the magnetically quenched system: at time $t=0$ the system is polarised along the z-axis, upon relaxation we report the time dependence of the magnetic precession at a temperature $T=100$K ($T=20$K) above (below) the Kondo temperature ($T_K\approx 37$K). The Fourier transform of the dynamics at $T=100$K shows a peak at $\omega=0.0029$eV ($=33.64$ K). The dynamics at $T=20$K shows sub-harmonics reminiscent at $\omega=0.0027$eV ($=31.32$ K) from the Zeeman splitting at time $t=0$.  
} \label{fig:kondo}
\end{center}
\end{figure}

\subsection*{Kondo physics} 
We extend further the application of MLDE to a realistic study case of the deposition of a correlated Kondo molecule on a gold surface (see Fig.\ref{fig:kondo}.a). Stable organic radical molecules exhibit a Kondo peak in the low-temperature experimental conductance, which is due to the presence of a single unpaired electron in the highest occupied molecular orbital \cite{ji.hi.13,zh.ka.13,frisenda2015kondo}. In the gas phase these molecules are paramagnetic. Due to their low spin-orbit interaction and small hyperfine splitting, they are expected to exhibit long spin-coherence times, and therefore have potential as building blocks of molecular spintronics applications \cite{frisenda2015kondo}. When brought in contact with a metal surface, the system corresponds to a single impurity Anderson model (SIAM) and has been modeled in the past using CTQMC or NRG as impurity solvers \cite{Droghetti.17,PTM_Kondo2018}. To demonstrate the capability of MLDE to describe Kondo physics, we choose the 1,3,5-triphenyl-6-oxoverdazyl (TOV) organic radical molecule, which, when deposited on a Au substrate has been shown experimentally to exhibit a Kondo temperature of about $\approx 37$K \cite{ji.hi.13}. Compared to other radical molecules on surfaces, this molecule has the advantage to have a well-defined contact geometry, where the molecule lies flat on the surface. We use the same simulation setup and parameters as in Ref. [\cite{Droghetti.17}] so that also the hybridization function of the SIAM is the same (see Fig. S3 in supplementary information). Describing Kondo physics at low temperature is a notoriously difficult problem for quantum impurity solvers. In particular, the collapse of energy scales in the Kondo limit prevents typical Lanczos or ED solvers from capturing the Kondo resonance, as the finite discretization tends
to introduce fictitious gaped states at very low temperature. 
We performed calculation at $T=5$K. In the Kondo regime (see Fig.\ref{fig:kondo}.b) the hybridization remains constant up to the lowest frequency. We note that $\Delta_1(i\omega_n)=\Delta_2(i\omega_n)$, which
confirms that the embedding source field $\delta \Delta(i\omega_n)$ remains negligible. 
Furthermore, MLDE fares better with $N_b=4$ than the best possible
fit obtained by the usual discretized Lanczos approach ($N_b=12$, triangles in
Fig.\ref{fig:kondo}.b), as the hybridization vanishes at small frequencies
(in the molecule, it remains constant). 
The imaginary part of the self-energy obtained by MLDE shows a Fermi liquid type behavior at small frequencies (see Fig.\ref{fig:kondo}.c), whereas the self-energy obtained by ED shows an artificial Mott singularity, due to the bath
discretization. We note that this low temperature was beyond the reach of our CTQMC solver ($\beta=35000$). Below the Kondo temperature, we recover the
Fermi liquid behavior of the self-energy (see Fig.\ref{fig:kondo}.d). 
As the MLDE representation is compact, it opens a large degree of possible
manipulation once the AIM is established. In particular, we extended the calculation to the time dynamics of the Kondo molecule with the Keldysh formalism \cite{keldysh_ed_lanczos} after a magnetic quench,
where at time \red{$t_0=0$} the molecule is magnetically polarized along the $e_z$ axis,
and the external magnetic field released for $t>t_0$. The magnetic moment
enters in a precession dynamics with the frequency equal to the Kondo temperature 
 (see Fig.\ref{fig:kondo}.e).\red{The estimated Kondo temperature which we achieved with different methods, e.g. from the time dynamics $(T_K\approx 33.64)$ and from perturbation theory\cite{Lee2019,Surer_2012} ($T_K\approx 33$, see Section 3 is supplementary information), which provides a further check of the validity of the MLDE approach.} The dynamics can also be resolved below $T_K$,
where  we observe additional harmonics, reminiscent from the Kondo Zeeman splitting at $t=t_0$ (see Ref.\cite{kondo_zeeman}).

\begin{figure}[!t]
\begin{center}
\hspace*{-0.5in}
\includegraphics[width=0.7\columnwidth]{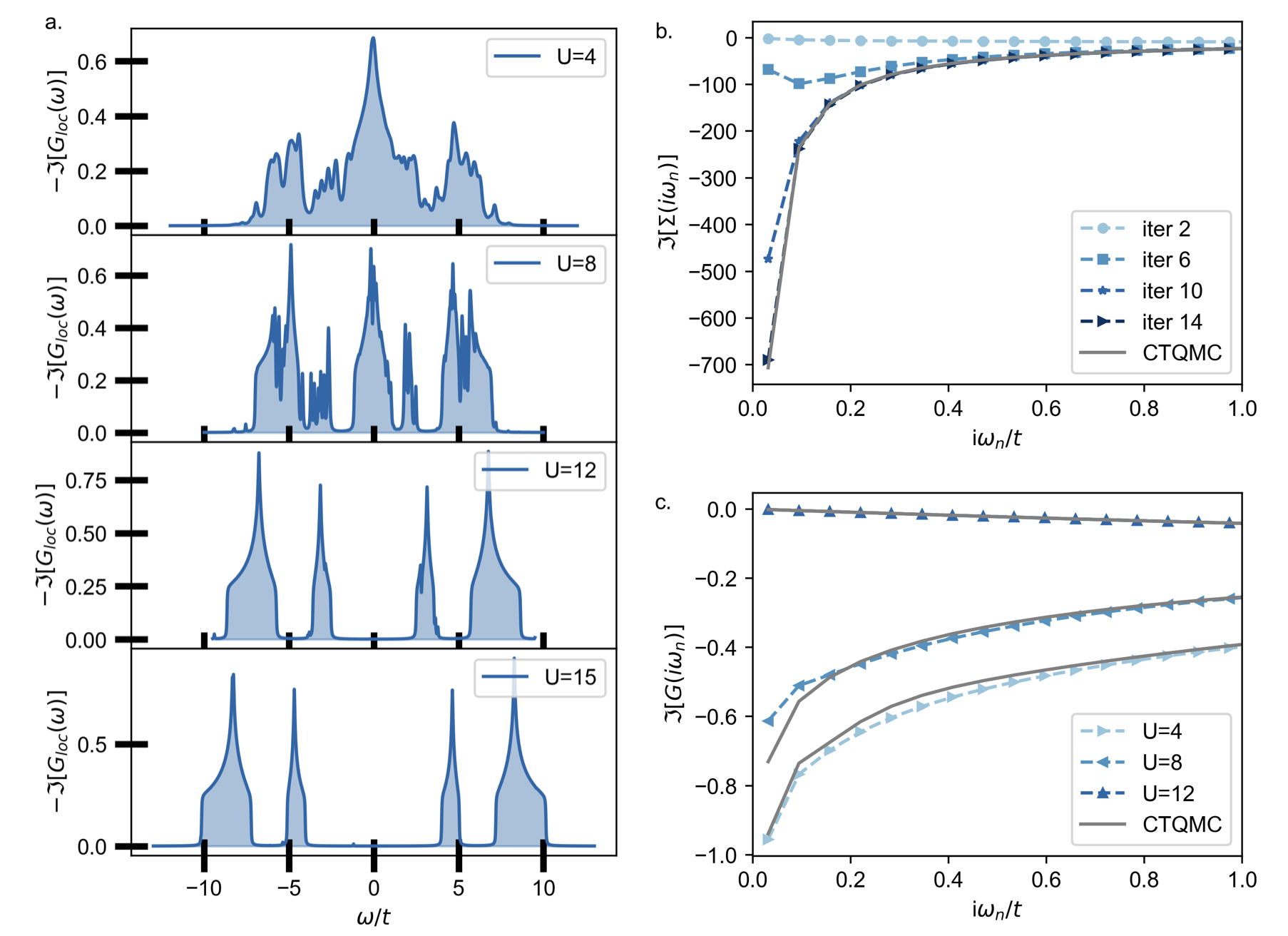}
\caption{\textbf{Mott transition}. (a) Spectral function obtained for the square lattice Hubbard
model solved by MLDE ($N_b=3$ for all calculations) with increasing
values of the on-site dimensionless correlation ratio $U/t$. For $U$ smaller
than the critical value $U_c \approx 9$, the system remains metallic, with a quasi-particle
weight (sharp peak at zero), Hubbard lower and upper bands, and band edge satellites that
develop near $U/t \approx 6$. In the Mott phase ($U/t>10$) we obtain a Mott charge gap $\Delta \approx U-W$, where $W/t=8$ is the bandwidth for the square lattice. (b) Imaginary part of the self-energy obtained for $U/t=12$ at different MLDE iteration of the self-consistent cycle. At iteration $14$, the self-energy is converged, and within the error bar or the converged CTQMC solution. (c) Converged imaginary part of the MLDE Green's function for different values of $U/t$ (symbols) compared with the DMFT CTQMC solution (dashed lines). Note the real part of the Green's function is zero in a particle-hole symmetric system.
} 
\label{fig:mott}
\end{center}
\end{figure}

\subsection*{Mott transition}
We have so far focused on simple AIM systems in the absence of mean-field corrections to the hybridization. We now turn to the dynamical mean-field MLDE approach, applied to the Mott transition \red{ where the local Green's function is defined with the density of state of} the two-dimensional square-lattice Hubbard model (see Fig.\ref{fig:mott}.a). Using the MLDE as a solver for DMFT, we recover the well-known metal-insulator transition (MIT) associated with the charge localization induced by the local Hubbard repulsion $U$.\footnote{\red{We clarify that DMFT self-consistency is enforced in Matsubara frequency.}} We recover with a simple MLDE and $N_b=3$ bath sites the spectral function of the Hubbard model, with the usual features (lower and upper Hubbard bands, quasi-particle peak below the transition $U_c \approx 9$, Mott gap for $U>U_c$). We note that interestingly the spectral function obtained by MLDE also shows the satellite peaks at the gap edge
for $U>U_c$. \red{This feature is typically difficult to obtain with analytically continued spectral function from Matsubara quantity. Indeed the inner peak near the Mott gap edge are associated with the real part of the self-energy, and in particular, is derived from an actual quasi-particle solution \cite{Granath_2014}.}\\
The MLDE charge gap $\Delta$ reproduces the known trend $\Delta=U-W$, which provides a further test of the theory. The benchmark with the converged solution obtained by CTQMC is remarkable (see Fig.\ref{fig:mott}.b), and the MLDE solution essentially within the error bars of the CTQMC for $U/t=12$. Across the Mott transition, the agreement between the Green's functions remains good (see Fig.\ref{fig:mott}.c). 

\begin{figure}[!t]
\begin{center}
\includegraphics[width=0.7\columnwidth]{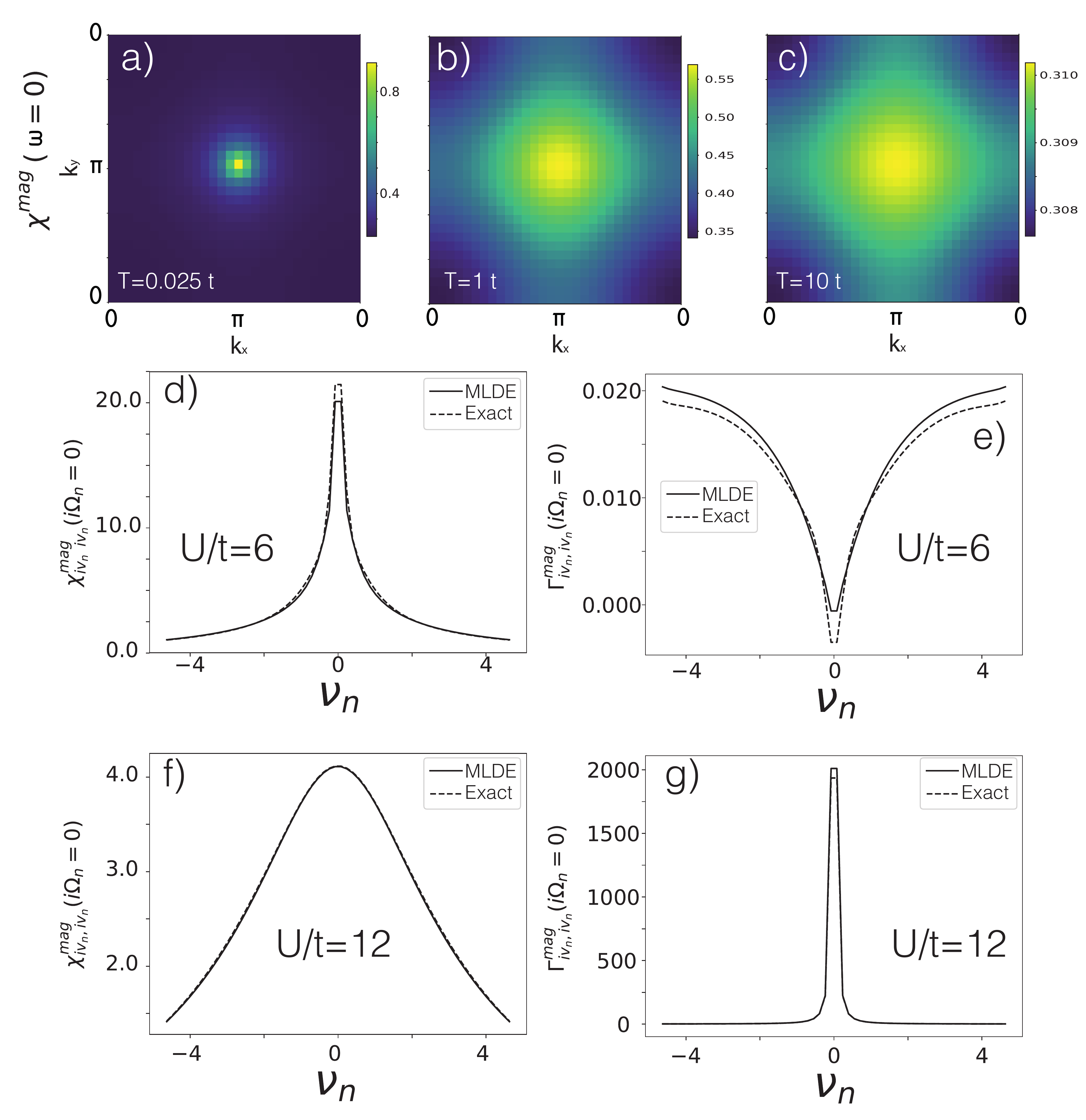}
\caption{\textbf{Dynamical susceptibility}. 
 Momentum resolved spin susceptibility obtained by Bethe Salpeter with the vertex calculated
with MLDE in the Hubbard model with $U/t=12$, at increasing temperature $T/t=0.025$ (a), $T/t=1$ (b), $T/t=10$ (c). We report that the magnetic susceptibility at $(\pi,\pi)$ is enhanced as the system
reaches the meltdown of the Mott gap (b), and then at very high temperature (c) becomes uniform. As the MLDE calculations only involves $N_b=3$ bath sites, the vertex is fully tractable at any temperature. 
d) Magnetic susceptibility $\chi$ and e) irreducible vertex $\Gamma$ resolved in fermionic frequency $i \nu$ obtained by MLDE (continuous line) and compared with the exact vertex (dashed line) at temperature $T/t=0.025$ and $U/t=6$. Respectively $\chi$ (f) and $\Gamma$ (g) obtained in the Mott phase for $U/t=12$. The agreement is remarkable.
} 
\label{fig:vertex}
\end{center}
\end{figure}

\subsection*{Bethe-Salpeter Equation}
We extended the calculations to the Bethe-Salpeter Equation (BSE) formalism,
applied to the MLDE solution. In particular we calculate the local irreducible
vertex $\Gamma$ (see supplementary information), which enables the calculation of the non-local dynamical magnetic susceptibility $\chi^{mag}(\omega)$ within MLDE. We performed calculations for $U/t=12$ at various temperatures, to explore the
behavior of magnetic excitations across the Mott gap melting. Within the
Mott gap phase at temperature $T/t=0.025$ (Fig.\ref{fig:vertex}.a), the 
Neel fluctuations are local in momentum at $Q_{\text{Neel}}=(\pi,\pi)$. As expected, the spin fluctuations are largely located at $Q_{\text{Neel}}$ in the Mott phase ($T=0.025 t$). \red{We observe that the antiferromagnetic magnetic fluctuations become gradually more incoherent as the temperature is increased throughout (Fig.\ref{fig:vertex}.b). Interestingly, we observe a large degree of incoherence at all considered temperature, which is reminiscent of the large scattering rate of the archetypal Mott system La2CuO4.}
 At very large  temperature (Fig.\ref{fig:vertex}.c), the fluctuations are fully incoherent, and the spectrum uniform across the Brillouin zone.

We note that a known challenge for the BSE approach is the
extraction of the local irreducible vertex $\Gamma$, which is obtained
by calculating the two-electron response function $G^{(2)}$. Such a quantity
requires traditionally computationally demanding collection of statistics by CTQMC and few other alternatives exist. The vertex can be calculated in the Lehman representation (see supplementary information), but requires sampling over the whole Hilbert space, which is not possible for AIM with more than few bath sites. In MLDE this task is largely simplified
as the Hilbert space remains compact. We provide a benchmark of the local vertex
$\Gamma$ with CTQMC for both $U/t=6$ and $U/t=12$ (see respectively 
Figs.\ref{fig:vertex}.e and \ref{fig:vertex}.g), in both cases the
agreement is quantitative. This agreement is also obtained in the fermionic representation of the local magnetic susceptibility $\chi_{i\nu,i\nu}$
(see Figs.\ref{fig:vertex}.d and \ref{fig:vertex}.f).


\begin{figure}[h]
  \centering
  \includegraphics[width=0.4\columnwidth]{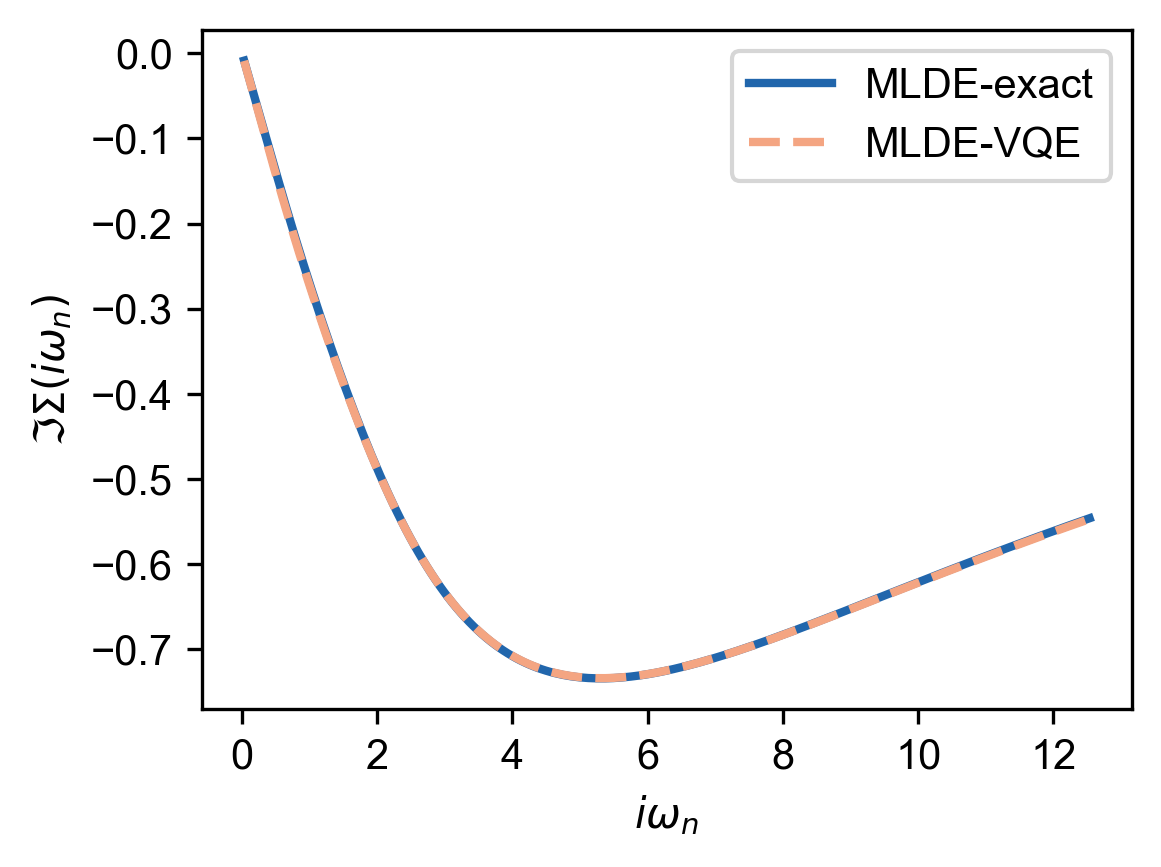}
  \caption{\textbf{MLDE on a quantum computer} Imaginary part of the self-energy as function of the Matsubara frequencies for the MLDE system presented in Fig. \ref{fig:aim} and $U=6$, computed using the quantum computing algorithm (``MLDE-VQE'', blue solid curve), and compared to the results obtained with the conventional computing algorithm used also for all results presented in Fig. \ref{fig:aim}  (``MLDE-exact'', red dashed curves).}
  \label{fig:mlde_qc}
\end{figure}

\subsection*{MLDE on a quantum computer}
On classical computers, the computational cost increases exponentially with system size, both for ED and MLDE, while quantum computers can potentially reduce this to a polynomial scaling \cite{PhysRevA.92.062318,PhysRevX.6.031045}. On near term quantum computers the noise limits the size of calculations that can be done in practice, so that only very small proof of concept DMFT systems have been demonstrated on hardware \cite{rungger2020dynamical}. Since MLDE significantly reduces the number of sites needed to describe a given physical system, it allows to reduce the required number of qubits as well as the depth of the quantum circuits. Therefore, in the near term, it is an ideal framework to run larger physical systems on NISQ devices, while in the long term, the reduction of the required number of sites will be greatly beneficial also in the fault-tolerant quantum computing regime.

Therefore, here we propose a quantum algorithm to run MLDE on NISQ devices and demonstrate it using a quantum simulator on the benchmark system considered in Fig. \ref{fig:aim}. To calculate the Green's function on a quantum computer, we use the variational quantum eigensolver (VQE) based method presented in Ref. \cite{rungger2020dynamical}, which was demonstrated to run on currently available hardware as it is rather resilient to noise. Details of the method are presented in the supplementary information. Here we show results obtained with our implementation within the Quest quantum simulator\cite{quest}. To demonstrate that the quantum MLDE algorithm gives the same accuracy as the classical MLDE algorithm, we apply it on the same Anderson impurity model shown in Fig. \ref{fig:aim}. The system has one impurity site and two bath sites, which we map to a 6 qubit system using a Jordan-Wigner transform \cite{PhysRevLett.63.322}. In Fig. \ref{fig:mlde_qc} we show that the self-energy for $U=6$ computed with the quantum computing algorithm agrees very very well with the result obtained using the classical computing algorithm. We have verified that the same is true for all values of $U$. This, therefore, demonstrates the functionality of our quantum algorithm for MLDE.

\begin{figure}[h]
  \centering
  \includegraphics[width=0.7\columnwidth]{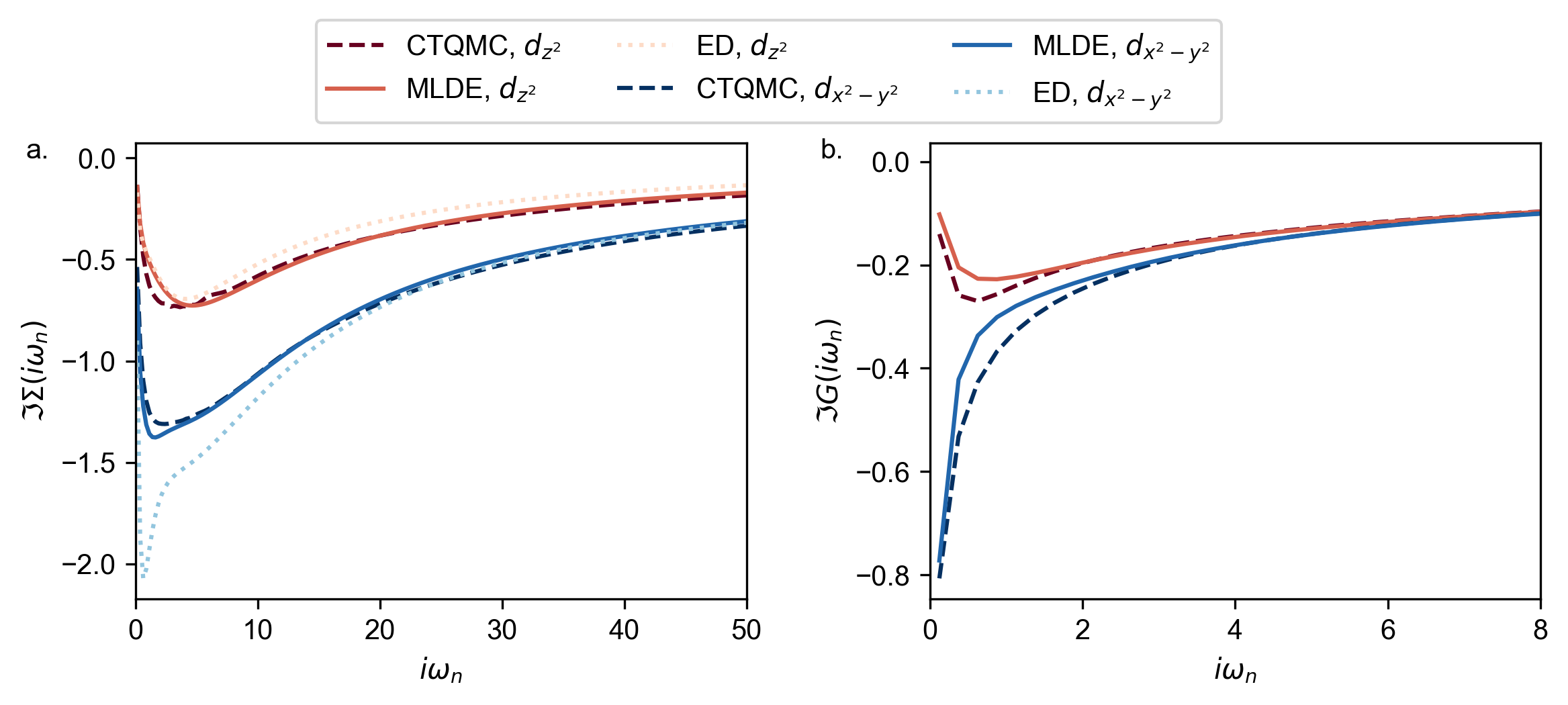}
  \caption{\textbf{MLDE multiorbital correlated LaNiO$_3$}:  We compare the Self energy of the first iteration of the self-consistent DMFT loop (panel a) and the Green's function for the final iteration (panel b) for the $e_g$ orbitals of the Ni ion in the LaNiO$_3$ ($d_{z^2}$ in red and $d_{x^2-y^2}$ in blue). In both case, we compare the quantities obtained with MLDE with 3 sites in the bath (solid line) with the quantities obtained with CTQMC (dashed line). In both case, the MLDE is in a good qualitative agreement with the CTQMC result. }
  \label{fig:multiorbital}
\end{figure}

\subsection*{MLDE for multiorbital system}
So far, we have considered a single orbital case as a proof of concept of the formalism. To highlight the novelty and the advantage of the MLDE solver, we now turn to the application to a multi-orbital case. 
The ED solver has been notoriously limited for the case of multi-orbital systems (see Ref. \cite{Liebsch_2011})as it would require a large number of bath sites, whereas CTQMC solver would be affected by the sign problem triggered by the non-zero off-diagonal terms of the hybridization between the orbitals.
Therefore, in this section, we demonstrate the ability to treat multi-orbital systems within MLDE framework.
As a test system, we applied DFT+DMFT on LaNiO$_3$, whose nontrivial contributions stemming from electronic correlations have been studied using CTQMC solver \cite{PhysRevB.92.245109}. In particular, its $e_g$ active sites pose this compound as our ideal candidate of two orbitals system. 
Our main focus is the comparison between a standard DFT+DMFT implementation using CTQMC and ED impurity solver with our MLDE environment with 3 bath sites. As shown in Fig 8a, a good agreement is achieved between CTQMC and MLDE when considering the local self energy for the first iteration of the DMFT loop.
We also show the local self energy obtained with ED using 3 bath sites. Notice that, differently from the results obtained by MLDE, the ED solver does not well capture the correct self energy behavior. A better agreement between the ED solver and the CTQMC would be achieved at the cost of adding more bath sites to increase the number of degrees of freedom for the fit of the hybridization as it is done in MLDE.\\
Lastly, we consider the local Green's function for the final iteration of the self-consistent DMFT loop. Notice that while we only include 3 sites in the bath, the MLDE Green's function is in a good agreement compared to CTQMC results. 

\section*{Discussion} 
In conclusion, we introduced a novel quantum many-body embedding techniques, the so-called \emph{Maximally Localized Dynamical Embedding} approach (MLDE), which offers a robust and efficient methodology to describe electronic correlations in quantum materials. This approach generalizes the local dynamical mean field theory to minimally non-local Anderson impurity model, remarkably opening up  the resolving power of the discretized impurity model. \red{In particular, when
the number of poles described by a canonical DMFT hybridization scales linearly with the number of quantum impurity sites, the latter increases exponentially in MLDE.} This opens new avenues in the realm of quantum computing where, with a realistic number of qubits (typically 10-20), MLDE would allow describing transition metal systems on a quantum computer, with minor errors induced by the bath discretization. As the MLDE hamiltonian is minimal (typically 4 sites in total), we have shown that time evolutions (Keldysh) and vertex calculations become routinely possible at a minimal cost. Finally, the correlated bath sites of MLDE also allows to describe Kondo physics at very low
temperature, which is a known limitation of standard exact diagonalization DMFT approaches, and it lays the foundations for future works on time-dependent non-equilibrium phenomena.

\section*{Methods}

Within DMFT, the effective AIM is subject to a self-consistency condition which relates the Green's function of the impurity model $G(i\omega_n)$ to the so-called Weiss-field ${\cal{G}}_0^{-1}(i\omega_n)$, which completely characterizes the AIM. For the single-band case, the Hamiltonian of the AIM reads:

\begin{equation}
\label{aim}
H=\sum\limits_{ij\sigma}{\epsilon_{ij\sigma} \hat d^\dagger_{i\sigma} \hat d_{j\sigma}} 
+\sum\limits_{i\sigma}{V_{i}( \hat d^\dagger_{i\sigma} \hat f_{\sigma}+hc)} + U \hat n_{f\uparrow} \hat n_{f\downarrow}  + \sum_{\sigma}\epsilon_f \hat f_{\sigma}^\dagger \hat f_{\sigma}
\end{equation}

where $d^\dagger_{p\sigma}$ ($d_{p\sigma}$) creates (destroys) a particle with spin $\sigma$ in the d-orbitals of the uncorrelated bath ($p \in [1,N_b]$) and $f^\dagger_{\sigma}$ ($f_{\sigma}$) creates (destroys) a spin $\sigma$ particle on the impurity, $U$ is the static Coulomb repulsion on the impurity and $V$ is the tunneling amplitude between the impurity and the bath. \\
In the maximally localized dynamical embedding approach (MLDE), we add an additional general two-electron interaction to the bath sites, and the interaction vertex reads:

\begin{equation}
H_{\text{int}} = \sum\limits_{\sigma_1,\sigma_2}{ \sum\limits_{i,j,k,l} {U^{(2)}_{i,j,k,l} \hat d^\dagger_{i\sigma_1} \hat d^\dagger_{j\sigma_2} \hat d_{k\sigma_2} \hat d_{l\sigma_1}}}
\end{equation}

The $U^{(2)}$ tensor is a fictitious two-body interaction between bath electrons introduced to enlarge the degrees of freedom of the approach. Although the free electron propagator takes a simple polynomial form, e.g. $G_0^{-1}=i\omega_n-\epsilon_{f}-\sum_i { V_i^2 / \left( i\omega_n - \epsilon_i \right)}$, a correlated green's function is instead described by a mapping to an exponentially long one-dimensional chain within the Kryvlov space \cite{green_moment}. In particular, when the bath is correlated, the Weiss field incorporates the dressed propagator of the bath electrons $G_\text{d}^{(1)}$: 

\begin{equation}
G_0^{-1}=i\omega_n-\epsilon_{f}-\underbrace{V^{\dagger}G_\text{d}^{(1)} V}_{\Delta^{(1)}},
\end{equation}

where the hybridization to the bath is denoted as $\Delta^{(1)}$. We emphasize that at this level of the theory, the self energy of the impurity $\Sigma_f$ remains naught, and hence we have a fully local theory from the perspective of the impurity. However, as we introduce a local correlation $U$ on the impurity, two effects occur: i) the propagator of the bath $G_d^{(1)}$ is dependent on the correlation on the impurity, and hence changes to $G_d^{(2)}$ and ii)  a non-local part of the self-energy $\Sigma_{fd}$ between the impurity and the bath emerges:

\begin{equation}
    G^{-1}=i\omega_n-\epsilon_{f}-\underbrace{(V^\dagger+\Sigma_{fd}^\dagger) G_\text{d}^{(2)} (V+\Sigma_{fd})}_{\Delta^{(2)}}-\Sigma_{f}
\end{equation}

We have now a set of embedding equations, which leads to a generalized Dyson equation $G^{-1}(z) - G_0^{-1}(z) = \Sigma_f(z) - \delta \Delta(z)$, where the $\delta \Delta$(z) is a source field that stems from the non-local correlations. Indeed, the latter term can be rationalized following a simple argument via the Migdal energy functional, which in MLDE reads

\begin{equation}
U \langle \hat n_\uparrow \hat n_\downarrow \rangle  = \frac{1}{2} \left( Tr \left( \Sigma_{f}G_{f} \right) + \underbrace{ Tr\left(\Sigma_{fd} G_{fd}\right)}_{\bar{Z}} \right) 
\end{equation}

Part of the correlation energy spills effectively on the bath \cite{Zgid_2017}, 
leading to a correlation leakage term $\bar{Z}$. The DMFT equations can be recovered when the leakage potential $\delta \Delta(z)$ and leakage correlation energy are small. As the tuning of the bath propagator allows for a very large set of parameters, these constraints can be successfully enforced via Lagrange parameters \red{( see cost function defined in Eq. 18 of supplementary information)} in the fit of the DMFT hybridization, to maximize the locality of the embedding, with a concomitant exponential improvement of the bath discretization errors. 
It is, however, known that in high dimension, or when a system is strongly correlated, the electron self-energy is well separable into a local dynamical part and a static non-local contribution \cite{selfseparable}. In this
respect, we reabsorb the embedding potential into a shift of the static part of the MLDE self energy, which ensures that the Migdal energy remains exact. \red{Further details on the derivation of the formalism are reported in the supplementary information.}

\section*{Acknowledgements}
CW acknowledges insightful and stimulating discussions with Andrew Mitchell.
CW was supported by grant EP/R02992X/1  from the UK Engineering and Physical Sciences Research Council (EPSRC). 
I. R. and  F. J. acknowledge the support of the UK government department for Business, Energy and Industrial Strategy through the UK national quantum technologies programme (InnovateUK ISCF QUANTIFI project). 
C.L. and F. J. are supported by the EPSRC Centre for Doctoral Training in Cross-Disciplinary Approaches to Non-Equilibrium Systems (CANES, EP/L015854/1). F.J is supported by  the Simons Many-Electron Collaboration. 
This work was performed using resources provided by the
ARCHER UK National Supercomputing Service and the Cambridge Service for Data Driven Discovery (CSD3)
operated by the University of Cambridge Research Computing Service (www.csd3.cam.ac.uk), provided by Dell EMC and Intel using Tier-2 funding from the Engineering and Physical Sciences Research Council (capital grant EP/P020259/1), and DiRAC funding from the Science and Technology Facilities Council (www.dirac.ac.uk). 

\section*{Author contributions statement}
CW designed the research. CL and FJ contributed to the code design. IR and FJ implemented the quantum computing algorithm.  All authors contributed to the results production and analysis. 
All authors reviewed the manuscript. 

\section*{Additional information}
The authors declare no competing interests.

\section*{Data and code availability}
The data and the code that support the findings of this study are available  at the  following repository  \url{ https://codeocean.com/capsule/2140807/tree/v2} or from the corresponding authors (carla.lupo@kcl.ac.uk, cedric.weber@kcl.ac.uk and ivan.rungger@npl.co.uk). 

\newpage

\setcounter{figure}{0}
\setcounter{equation}{0}
\renewcommand{\thefigure}{S\arabic{figure}}
\section*{Supplementary information}



\section{MLDE derivation}
In the manuscript associated to this supplementary information, we introduce a new formalism where the bath is represented by interacting electrons ('dynamical embedding'), rather than by free electrons as in other ED methods. This addition allows for an increase in the number of fitting parameters without increasing the number of bath sites. Considering an interacting bath comes at the cost of introducing non-local components of the impurity self-energy, which do not arise in DMFT, in which instead of the self-energy only depends on the bath via its hybridization function. To recover the correct self-energy, those non-local components are minimized or, in other words, 'Maximally Localised'.\\

In this section, we provide an extended derivation of the MLDE formalism. To this end, we start with reviewing the case of non-correlated bath. We then proceed to introduce the quantities of interest in the MLDE formalism where bath interactions are included. We also provide the pseudo-code where the algorithm is described in more detail. 
\subsection{Case of non correlated bath}
Within DMFT, the extended lattice model is mapped to an effective Anderson impurity model (AIM), where a correlated atom is embedded in a non-interacting bath.
As shown in Fig.1.a (main text), the bath represents the effective medium, and it is coupled to the impurity atom. Indeed electrons may hop from the atom to the bath with probability amplitude $V$ giving rise to quantum fluctuations of the local configuration on the single impurity. Within this mean-field approximation, the initial complex many-body problem is reduced to a local problem of a correlated single site where impurity-solver algorithms are applied.
We first introduce the Hamiltonian of the Anderson impurity problem:
\begin{equation}
    H=\sum\limits_{ij\sigma}{\epsilon_{ij\sigma} \hat d^\dagger_{i\sigma} \hat d_{j\sigma}} 
+\sum\limits_{i\sigma}{V_{i}( \hat d^\dagger_{i\sigma} \hat f_{\sigma}+hc)} + U \hat n_{f\uparrow} \hat n_{f\downarrow}  + \sum_{\sigma}\epsilon_f \hat f_{\sigma}^\dagger \hat f_{\sigma}
\label{eq:Hmlde}
\end{equation}
where $d^\dagger_{p\sigma}$ ($d_{p\sigma}$) creates (destroys) a particle with spin $\sigma$ in the $d$-orbitals of the uncorrelated bath ($p \in [1,N_b]$) and $f^\dagger_{\sigma}$ ($f_{\sigma}$) creates (destroys) a spin $\sigma$ particle on the impurity, $U$ is the static Coulomb repulsion on the impurity and $V_i$ is the tunnelling amplitude between the impurity and the bath.
The free-electron propagator is 
\begin{equation}
G_0^{-1}(i\omega_n)=i\omega_n-\epsilon_f-\Delta(i\omega_n)
\end{equation}
In the exact diagonalization (ED) approximation, the continuum bath is represented with a finite number of effective sites.
Hereby, we consider and AIM in a start geometry, where the hybridization $\Delta(i\omega_n)$ is discretized as follows:
\begin{equation}
    \Delta(i\omega_n)=\sum_d\frac{V_d}{i\omega_n-\epsilon_d}  
\end{equation}\label{eq:free_hybrid_non_interacting_bath}  
In the case of correlated impurity, $U\neq0$, the problems acquire a many-body nature, and the electron propagation is affected by the scattering events, therefore correlations between the electrons in the system.  We then introduce the interacting propagator of the impurity of the AIM in the function of the hopping terms $V_d$ between impurity and bath sites and onsite energies $\epsilon_d$ of the bath sites: 
\begin{equation}
G^{-1}_{imp}(i\omega_n)=i\omega_n-\epsilon_f-\Delta(i\omega_n)-\Sigma(i\omega_n)
\end{equation}
The Dyson equation for the impurity reads:
\begin{equation}
G^{-1}_{imp}(z)=G^{-1}_{0}(z)-\Sigma_f (z)    
\end{equation}
For our convenience, we report the Green's function and the self-energy $\Sigma$ of the problem, in a matrix formalism  Hence: 
\begin{equation}
\boldsymbol{\Sigma}=
\left(
\begin{array}{c|c}
   \Sigma_\text{f} & 0\\
   \hline
   0 & 0\\
\end{array}
\right)
\quad \text{and} \quad
\boldsymbol{G}^{-1}=
\left(
\begin{array}{c|c}
   (i\omega_n-\epsilon_{f}-\Sigma_{f}) & -\mathbf{V}\\
   \hline
   -\mathbf{V}& (i\omega_n-\mathbf{\epsilon}_{d})\\
\end{array}
\right)
\end{equation}
Notice that their dimension is $N_b+1$, with $N_b$ being the number of bath sites.
\subsection{Case of correlated bath}
In the maximally localized dynamical embedding approach (MLDE), 
we extend the formalism of the ED approximation to include correlations in the bath sites, adding a two-body interaction term in the Hamiltonian, which reads:
\begin{equation}
    H=\sum\limits_{ij\sigma}{\epsilon_{ij\sigma} \hat d^\dagger_{i\sigma} \hat d_{j\sigma}} 
+\sum\limits_{i\sigma}{V_{i}( \hat d^\dagger_{i\sigma} \hat f_{\sigma}+hc)} + U \hat n_{f\uparrow} \hat n_{f\downarrow}  + \sum_{\sigma}\epsilon_f \hat f_{\sigma}^\dagger \hat f_{\sigma}+\sum\limits_{\sigma_1,\sigma_2}{ \sum\limits_{i,j,k,l} {U^{(2)}_{i,j,k,l} \hat d^\dagger_{i\sigma_1} \hat d^\dagger_{j\sigma_2} \hat d_{k\sigma_2} \hat d_{l\sigma_1}}}
\end{equation}
The $U^{(2)}$ tensor is a fictitious two-body interaction between bath electrons, which has $N_b^2+N_b^2(N_b^2-1)/2$ nontrivial components where $N_b$ is the number of bath sites. Therefore the number of fitting parameters for the bath at most scales as $\approx N_b^4$ due to the presence of the two-body interaction.  
However, we want to emphasize that this formalism can be extended further by the addition of higher order terms (e.g. if a three-rank tensor is introduced, the number of parameters scales at most as $N_b^6$) being limited only by the dimension of the Hilbert space.

Consistent with the previous section, we start from the definition of the free-electron propagator. 
\begin{equation}
G_0^{-1}=i\omega_n-\epsilon_{f}-\underbrace{\boldsymbol{V}^{\dagger}\boldsymbol{G}_\text{d}^{(1)} \boldsymbol{V}}_{\boldsymbol{\Delta}^{(1)}},
\end{equation}
where the hybridisation to the bath is denoted as $\Delta^{(1)}$ and includes a Green's function for the bath $G_\text{d}^{(1)}$ which is interacting rather then being a free-electron propagator as in Eq.(\ref{eq:free_hybrid_non_interacting_bath}). Notice that at this level of the theory, the self-energy of the impurity $\Sigma_f$ remains naught, and hence we have a fully local theory from the perspective of the impurity.\\
We now turn to the case of non-zero correlation on the impurity. As stated in the main text, two effects occur:
\begin{itemize}
    \item the interacting propagator on the bath is now affected by the correlation in the impurity. Therefore a new definition $G_\text{d}^{(2)}$ is introduced 
    \item a non-local part of the self-energy $\Sigma_{fd}$ between the impurity and the bath emerges
\end{itemize}
Therefore the interacting Green's function is defined as 
\begin{equation}
    G^{-1}_{imp}=i\omega_n-\epsilon_{f}-\underbrace{(\mathbf{V}^\dagger+\boldsymbol{\Sigma}_{fd}^\dagger) \mathbf{G}_\text{d}^{(2)} (\boldsymbol{V}+\boldsymbol{\Sigma}_{fd})}_{\boldsymbol{\Delta}^{(2)}}-\boldsymbol{\Sigma}_{f}
\end{equation}
We have now a set of embedding equations, which leads to a generalized Dyson equation 
\begin{equation}
    \boldsymbol{G}^{-1}- \boldsymbol{G}_0^{-1} = \delta \boldsymbol{\Delta}-\boldsymbol{\Sigma}_{f}
\end{equation}
where the $\delta \Delta =\Delta_1-\Delta_2$ 
is a source field that stems from the non-local correlations.\\

To better clarify the formalism introduced so far, we recall the Green's function and the self-energy in their matrix form. Hence:

\begin{equation}
\boldsymbol{G}(i\omega_n)
=\left(
\begin{array}{c|c}
   G_\text{f}(i\omega_n) & \boldsymbol{G}_\text{fd}(i\omega_n)\\
   \hline
   \boldsymbol{G}_\text{df}(i\omega_n) & \boldsymbol{G}_\text{d}(i\omega_n)\\
\end{array}
\right)
\quad \text{and}\quad 
\boldsymbol{\Sigma}(i\omega_n)=
\left(
\begin{array}{c|c}
   \Sigma_\text{f}(i\omega_n) & \boldsymbol{\Sigma}_\text{fd}(i\omega_n)\\
   \hline
   \boldsymbol{\Sigma}_\text{df}(i\omega_n) & \boldsymbol{\Sigma}_\text{d}(i\omega_n)\\
\end{array}
\right)
\end{equation}
\footnote{Notice that to simplify the notation, the repeated indices are dropped. Therefore $\boldsymbol G_{dd}=\boldsymbol G_d$.}
A comparison with the matrices defined in the case of non-correlated bath, provides further clarification on the consequences of introducing interactions in the bath sites. 
The block of the self-energy matrix related to the bath has now an non-zero contribution by definition of our model. A less trivial observation concern the off-diagonal terms which are non zero, due to the interaction leaked in the bonds between impurity and bath sites. 
Therefore we can distinguish different contribution to the correlation energy, and in terms of Migdal energy functional, it reads:
  \begin{align}
        E_U^{tot}&=\frac{2}{\beta}\sum_{i\omega_n}\Tr{\left[\boldsymbol{G}^{\uparrow}(i\omega_n)\left(\boldsymbol{\Sigma}^\uparrow(i\omega_n)-\boldsymbol{\Sigma}^\uparrow(\infty)\right)\right]}+\Tr{\left(\boldsymbol{R}^\uparrow \boldsymbol{\Sigma}^\uparrow(\infty) \right)}\\
        E_U^{bath}&=\frac{2}{\beta}\sum_{i\omega_n}\Tr{\left[\boldsymbol{G}_{d}^{\uparrow}(i\omega_n)\left(\boldsymbol{\Sigma}^\uparrow(i\omega_n)-\boldsymbol{\Sigma}^\uparrow(\infty)\right)_{d}\right]}+\Tr{\left(\boldsymbol{R}_d^\uparrow \boldsymbol{\Sigma}_{d}^\uparrow(\infty) \right)}\\
        E_U^{int}&=\frac{2}{\beta}\sum_{i\omega_n}\Tr{\left[\boldsymbol{G}_{fd}^{\uparrow}(i\omega_n)\left(\boldsymbol{\Sigma}^\uparrow(i\omega_n)-\boldsymbol{\Sigma}^\uparrow(\infty)\right)_{df}\right]}\\
    E_U^{impurity}&=\frac{2}{\beta}\sum_{i\omega_n}G^\uparrow_{f}(i\omega_n)\left(\Sigma_{f}^\uparrow(i\omega_n)-\Sigma_{f}^\uparrow(\infty)\right)+R_{f}^\uparrow \Sigma_{f}^\uparrow(\infty)
    \end{align}
where $ R$ is the density matrix.
A crucial quantity of the formalism is the leakage field, which is defined both statically $E_U^{int}$, and dynamically  
\begin{equation}
    z_U^{int}(i\omega_n)=\frac{\left(\boldsymbol\Sigma_{fd}(i\omega_n)\cdot\boldsymbol G_{fd}(i\omega_n)\right)^{2}}{\vert i\omega_n\vert}
\end{equation}



Part of the correlation energy spills effectively on the bath \cite{embedding_field}, leading to a correlation leakage term, which is defined by 
\begin{equation}
    E_U^{leakage}=\frac{2}{R_{ff}}
    \left(E_U^{tot}-E_U^{bath}-E_U^{imp}\right)
\end{equation}

The locality principle behind the DMFT equations can be recovered when the leakage potential $\delta \Delta(z)$ and leakage correlation energy are small. 
As the tuning of the bath propagator allows for a very large set of parameters, these constraints can be successfully enforced via Lagrange parameters in the fit of the DMFT hybridization, to maximize the locality of the embedding, with a concomitant exponential improvement of the bath discretization errors. To that end, the cost function of the minimization procedure is defined as: 
\begin{equation}
  d=\sum_{i\omega_n}\frac{\vert \Delta_1-\Delta_{target}\vert^{2} +\vert \Delta_2-\Delta_{target}\vert^{2}+\alpha_3\vert \Delta_1-\Delta_{2}\vert^{2}}{\vert i\omega_n\vert^{\alpha_1}}+\lambda_s \vert E_U^{leakage}\vert +\lambda_d \sum_{i\omega_n} \vert z_U^{int}(i\omega_n)\vert 
\end{equation}

where the parameters $\alpha_1,\alpha_3,\lambda_d,\lambda_s$ are properly tuned according to the case of study to improve the minimisation procedure. We noticed that in most of the cases, the leakage field turns out to be naturally small. Indeed this follows from the constraint that $\vert \Delta_1-\Delta_2\vert $ has to be minimal. Therefore the fitting process tries to reduce the leakage by construction.

It is, however, known that in high dimension, or when a system is strongly correlated, the electron self-energy is well separable into a local dynamical part and a static non-local contribution \cite{selfseparable}. In this
respect, we reabsorb the embedding potential into a shift of the static part of the MLDE self-energy, which ensures that the Migdal energy remains exact. 

To summarise, in the MLDE formalism, non-local correlations are introduced to enlarge the parameters for the fitting of the parameters. This represents an improvement over existent ED formalism , as the increased number of fitting parameters allow the use of less bath sites. The non-locality is kept small through a Lagrange parameter in the cost function.  To recover consistency with MLDE, the leaked (non-local) correlated energy is reabsorbed in the impurity self-energy. More details on the different steps of the algorithm are shown in the pseudo-code (\textbf{Algorithm 1}) reported below.

\begin{algorithm}
\caption{MLDE solver}
\begin{algorithmic} 
\STATE inputs: $\Delta_{target},\boldsymbol\epsilon,\alpha,\alpha_3,\lambda_s,\lambda_d$, $U$, $N_b$ 
\STATE
\REQUIRE $U_{imp}=0 \vee U_{bath}\neq 0$ 
\STATE \textbf{call} ED solver: input: $\boldsymbol\epsilon$, $\mathbf{V}$, output $\boldsymbol\Sigma,\boldsymbol G$
\STATE $\boldsymbol\Sigma_{int}\leftarrow -\boldsymbol V-\boldsymbol\Sigma_{fd}$
\STATE $G_d^{(1)}\leftarrow -\boldsymbol\epsilon_b-\boldsymbol\Sigma_d+i\omega_n$
\STATE $\boldsymbol\Delta_1 \leftarrow \boldsymbol\Sigma_{int}^\dagger (\boldsymbol G_d^{(1)})^{-1} \boldsymbol\Sigma_{int}$
\STATE $d_1 \leftarrow \sum_n \frac{\vert\Delta_1-\Delta_{target}\vert^2}{\vert i\omega_n\vert^\alpha}$
\STATE
\REQUIRE $U_{imp}\neq 0 \vee U_{bath}\neq 0$ 
\STATE \textbf{call} ED solver: output $\boldsymbol\Sigma,\boldsymbol G$
\STATE $\boldsymbol\Sigma_{int}\leftarrow -\boldsymbol V-\boldsymbol\Sigma_{fd}$
\STATE $\boldsymbol G_d^{(2)}\leftarrow -\boldsymbol\epsilon_b-\boldsymbol\Sigma_d+i\omega_n$
\STATE $\boldsymbol\Delta_2 \leftarrow \boldsymbol\Sigma_{int}^\dagger (\boldsymbol G_d^{(2)})^{-1} \boldsymbol\Sigma_{int}$
\STATE compute $E_U^{tot},E_U^{bath},E_U^{imp}$ and $E_U^{leakage}$
\STATE $d_2 \leftarrow \sum_n \frac{\vert\Delta_2-\Delta_{target}\vert^2}{\vert i\omega_n\vert ^\alpha}$ 
\STATE $d_3 \leftarrow \sum_n \frac{\vert\Delta_2-\Delta_1\vert^2}{\vert i\omega_n\vert^\alpha}$ 
\STATE $d_4\leftarrow \vert E_U^{leakage}\vert$
\STATE $d_5\leftarrow \sum_{i\omega_n} \vert z_U^{int}(i\omega_n)\vert$
\STATE $d=d_1+d_2+\alpha_3 d_3+d_4\lambda_s+d_5\lambda_d$
\STATE $\Sigma_f\leftarrow \Sigma_f+E_U^{leakage}$
\end{algorithmic}
\end{algorithm}

\subsection{Benchmark in real frequency}
In the main text, we preferred to benchmark the MLDE with ED and CTQMC,  in the Matsubara formalism, rather than in real frequency. Indeed, the spectral functions are notoriously known for being a poor benchmark when using the continuous fraction method, used for the benchmark ED solver (as the number of bath sites is large, Lehmann representation of the GF is not achievable). In particular, reliable benchmarks can be obtained via the Keldysh formalism and Fourier transforming the time evolution to real frequency \cite{Wolf_2015}, but require extensive computing resources for the large systems used as the benchmark. In this regard, we compared the spectral functions obtained by the MLDE for the Mott transitions with the one obtained by ED (see Fig S2). However, a comparison in the real axis is less reliable due to the continuous fraction. Moreover, the ill-conditioned maxent approach prevents us from detailed comparison on the real axis between CTMQC and MLDE. Therefore in the main text of the manuscript, we have benchmarked the different solvers in the Matsubara formalism.


{\color{black}{
\section{Vertex calculations in MLDE}
In this section, we provide further insights on the application of MLDE to the calculation of two particle Green's function quantities. 
In particular, the dynamical susceptibility fermionic matrix is computed by inverting the Bethe-Salpeter Equation (BSE)
\begin{equation}
  \chi(q,i\Omega_n)_{(i\nu\sigma)(i\nu'\sigma')} = [(\chi^0(q,i\Omega_n)^{-1} + \Gamma^{imp}(i\Omega_n)]^{-1}_{(i\nu\sigma)(i\nu'\sigma')},
\end{equation}
where :
\begin{equation}
\chi(q,i\Omega)^0_{(i\nu\sigma)(i\nu'\sigma')} = -\delta_{i\nu i\nu'}\d_{\s\s'}\sum_kg_{k\s}(i\nu)g_{k+q,\s'}(i\nu+i\Omega) ,
\end{equation}
and $\Gamma^{imp}$  is the impurity vertex obtained by inverting the BSE for the MLDE impurity problem. 
The impurity susceptibility reads: 
\begin{align}
  G^{(2)imp}_{(i\nu\sigma)(i\nu'\sigma')}(i\Omega) &= \int_0^{\beta}d\tau_1 d\tau_2d\tau_3d\tau_4e^{i\nu(\tau_1-\tau_2)+i\nu'(\tau_3-\tau_4) + i\Omega(\tau_1-\tau_4)}                                                     \left<\mathcal{T}_{\tau}c_{\sigma}^{\dagger}(\tau_1) c_{\sigma}(\tau_2)c_{\sigma'}^{\dagger}(\tau_3) c_{\sigma'}(\tau_4)\right>\\
  \chi^{imp}_{(i\nu\sigma)(i\nu'\sigma')}(i\Omega) &= G^{(2)imp}_{(i\nu\sigma)(i\nu'\sigma')}(i\Omega) - \beta \delta_{0 \Omega}G(i\nu)G(i\nu').
\end{align}

In our work, we compute the two-particle response function $G^2$ via the Lehman representation\cite{vertex1,vertex2}: 
\begin{flalign}
G^{(2)imp}_{i\nu i\nu'}(i\Omega) &= \frac{1}{Z} \sum_{ijkl}\sum_{\Pi} \Phi(E_i,E_j,E_k,E_l;\omega_{\Pi_1},\omega_{\Pi_2},\omega_{\Pi_3})\text{sign}(\Pi)  \left< i|\mathcal{O}_{\Pi_1} |j\right> \left< j|\mathcal{O}_{\Pi_2} |k\right> \left< k|\mathcal{O}_{\Pi_3} |l\right> \left< l|c_{\sigma} |i\right> \\
\Phi(E_i,E_j,E_k,E_l;\omega_{1},\omega_{2},\omega_{3}) &= \int_0^{\beta} d\tau_1 \int_0^{\tau_1} d\tau_2\int_0^{\tau_2} d\tau_3e^{-\beta E_i +\tau_1(E_j-E_i)+\tau_2(E_k-E_j)+\tau_3(E_l-E_k)}e^{i(\omega_1\tau_1+\omega_2\tau_2+\omega_3\tau_3)},
\end{flalign}
where we have introduced the closure relation, and define $O_1=c_\sigma$,
$O_2=c^\dagger_\sigma$ and $O_3=c_{\sigma'}$.
$E_i$ are the eigenvalues of the MLDE hamiltonian, $\Pi$ are the triplet permutations, and the summation holds over all states of the Hilbert space.}}

The magnetic susceptibility is obtained as
\begin{equation}
    \chi^{mag}(q,i\Omega_n) = \frac{1}{2} \sum_{\s}\chi(q,i\Omega_n)_{(i\nu,\sigma)(i\nu',\sigma)} - \chi(q,i\Omega_n)_{(i\nu,\sigma)(i\nu',-\sigma)} .
\end{equation}

In principle, the minimization of the single particle leakage does not guarantee the minimization of the leakage for the two particle irreducible vertex. For the case we have studied, even without imposing a constraint on the two particle leakage, the agreement  with CTQMC is remarkable.  We propose here a diagrammatic argument to support our findings, in particular, that MLDE also provides a good estimate of higher order response functions.
        
        Firstly, the lowest order diagram contributing to the irreducible vertex is given by (see Ref. \cite{PhysRevB.86.125114}):
        \begin{equation}
            \Gamma^{(1)}_{abcd} = U_{abcd}.
        \end{equation}
        Since in MLDE, there are no direct many-body interaction terms in the Hamiltonian between the impurity and the bath, this diagram is equal to zero by construction for the cross term contribution. Furthermore, the MLDE non-local self-energy connecting the bath and the impurity can be expressed in term of the reducible vertex\cite{RevModPhys.90.025003}:
        \begin{equation}
        \Sigma(14) = -U(12'3'1')G(1'4')G(23')G(2'3)F(4'234)
        \label{eq:F_sigma}
        \end{equation}
        where $F$ is the reducible vertex, which is related to the irreducible vertex by the Bethe-Salpeter equation
        \begin{equation}
            F(1234) = \Gamma(1234) + \Gamma(122'3') \chi^0(3'2'4'3') F(3'4'34).
            \label{eq:F_BSE}
        \end{equation}
        Here we use the usual short-handed notations  $1=(\tau_1,l_1,\sigma_1)$ and assume sum over repeated indices. We now consider the case of  $\Sigma(14)$ 
        corresponding to the non-local
        self-energy that connects the impurity to the bath degrees of freedom. 
        When the left-hand side of eq \ref{eq:F_sigma} is minimized in MLDE (reduction of the one-particle leakage), it naturally imposes that $F$ remains small (for non-local contributions). Moreover, we see from eq \ref{eq:F_BSE} that minimizing $F$ also imposes a minimization of the non-local terms in $\Gamma$, that connects the impurity to the bath. In summary, when the non-local self-energy is minimal, MLDE provides a good approximation of the irreducible vertex used throughout to calculate the dynamical magnetic susceptibility. 
    
\carla{Finally, we emphasize that it is usually difficult to provide energy cutoffs on the low energy converged states of the Hamiltonian (typically in Lanczos or other iterative approaches), as the Boltzmann statistics only truncates one of the sums (on index $i$),
whereas all eigenstates are required for the other summation. MLDE provides an ideal candidate for such calculations, as the number of bath sites can be kept small with very little cost in accuracy. 
We note indeed that the previous equation can be arranged to show a computational complexity growing with the cube of the number of eigenstates $n$, where $n$ itself grows exponentially with the number of bath sites.}

\begin{figure}[!t]
\begin{center}
\includegraphics[width=0.5\columnwidth]{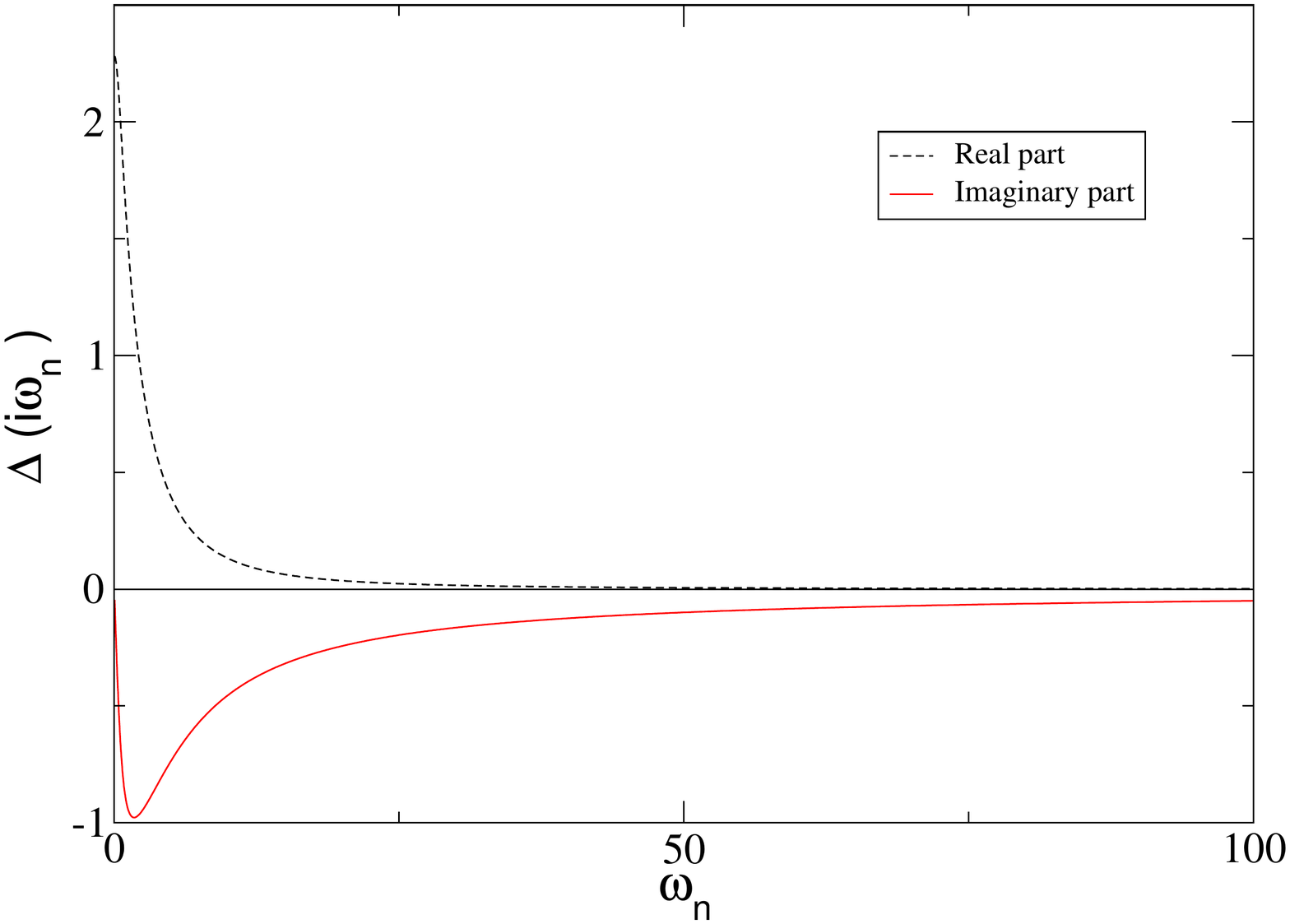}
\caption{\textbf{Model hybridization}. 
 Real part and imaginary part of the model hybridization used for the MLDE benchmark.} 
\end{center}
\end{figure}

\begin{figure}[!t]
\begin{center}
\includegraphics[width=1\columnwidth]{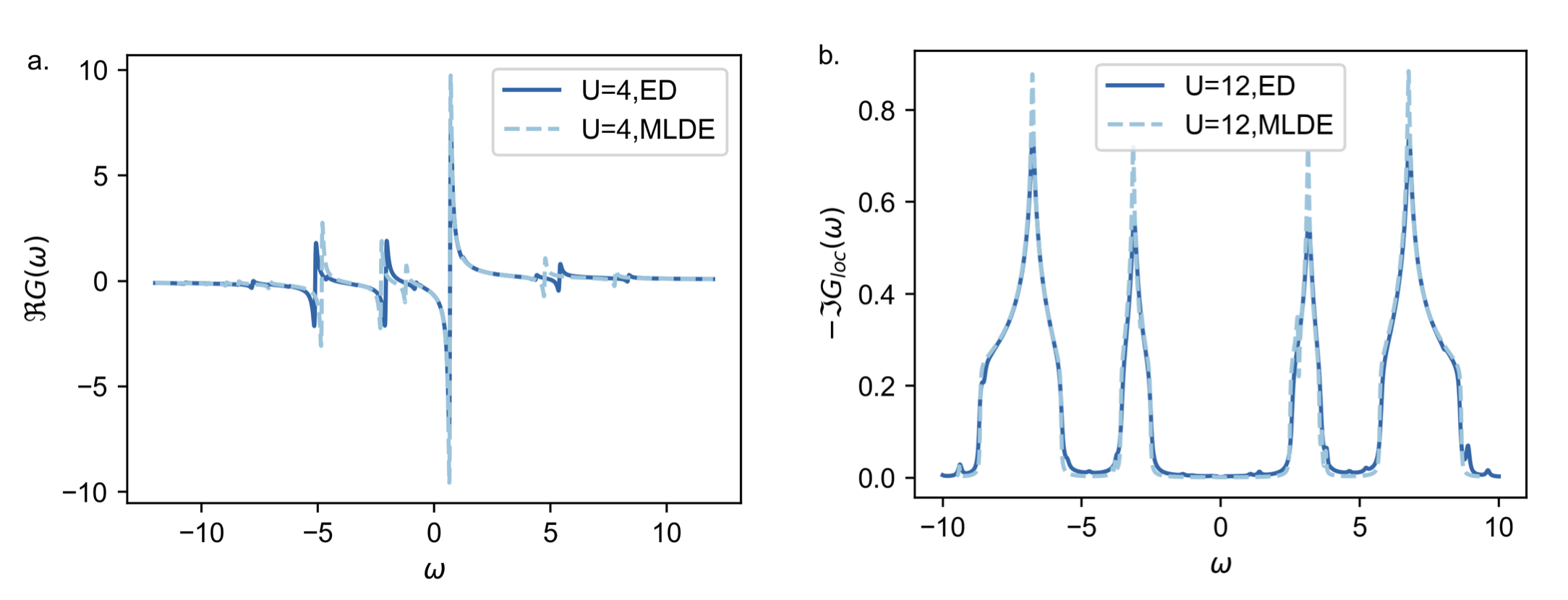}
\caption{\textbf{Benchmark between ED and MLDE in real-frequency} We compare the results obtained with ED (9 bath sites) and MLDE (3 bath sites). In panel a) we report the real par of the impurity Green's function obtained solving the AIM for fixed $U=4$. In panel b) we report the local Green's function obtained solving the self-consistent DMFT loop in the case of the square lattice, for fixed value $U=12$. } 
\end{center}
\end{figure}

\section{MLDE for Kondo molecule}

The density of states (DOS) for different temperatures is shown in Fig S4a. The fingerprint Kondo peak around zero energy can be seen at low temperatures. We emphasize that this feature is notoriously hard to obtain when using an ED solver as the formalism is affected by the small broadening factor needed to correctly capture this Kondo physics fingerprint. 
The Kondo temperature can be obtained from the relation\cite{Lee2019,Surer_2012} obtained from perturbation theory, which reads
\begin{equation}
    T_K=-\frac{\pi Z}{4}\Im\Delta(i\omega_n)\vert_{\omega_n\to 0}
\end{equation}
With the computed quasi-particle weight ($Z=0.026$) and value of the hybridization in the limit of zero frequency (see Fig S3), we obtain a Kondo temperature of $T=33.8$ K, which is comparable with the experimental one at $\approx 37$K \cite{ji.hi.13}.
We notice that temperature effect corrections to the Fermi liquid regime can be more easily studied in the Matsubara self-energy (see Fig S4.b). In this representation, the quadratic behavior of the self-energy is apparent at low energy scales.

\begin{figure}[!t]
\begin{center}
\includegraphics[width=0.5\columnwidth]{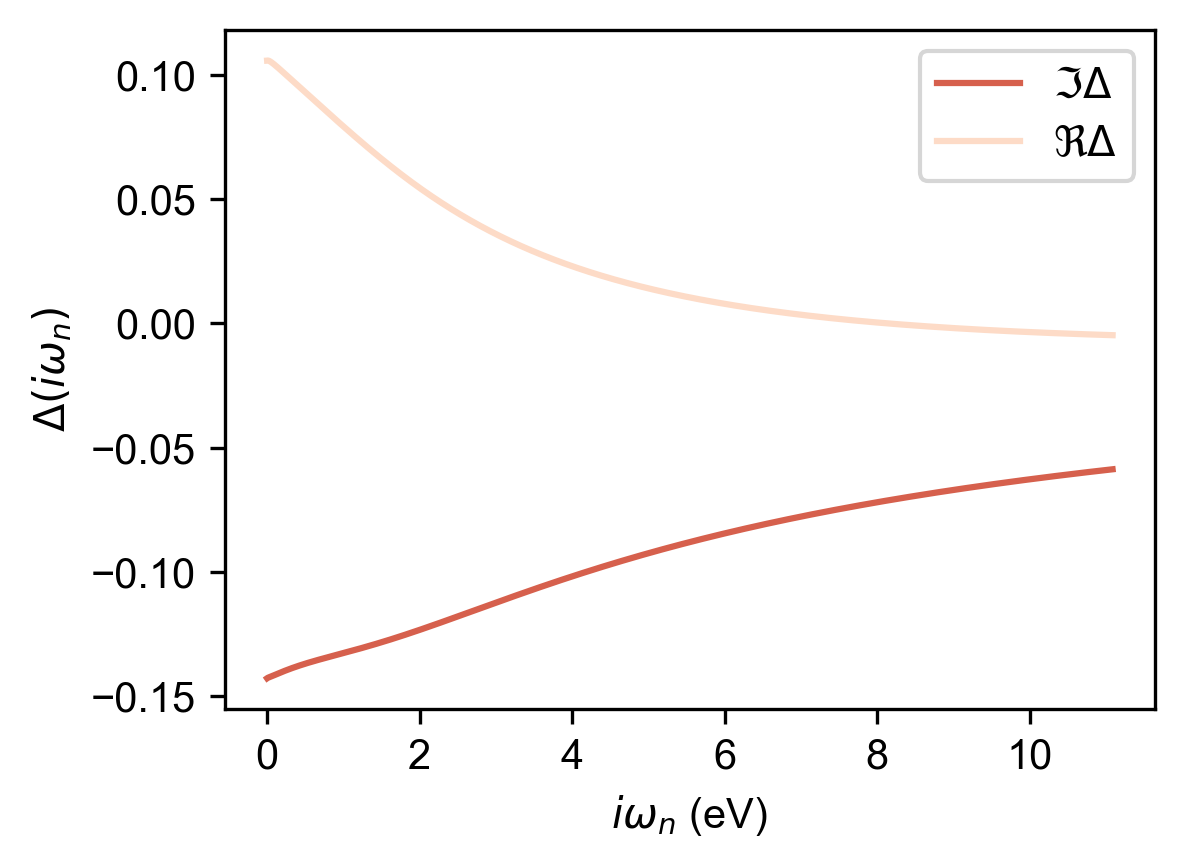}
\caption{\textbf{Model hybridization of the TOV molecule on Gold surface}. 
 Real part and imaginary part of the hybridization used for the application of MLDE to study the case of the deposition of the TOV molecule on gold surface as obtained in Ref.\cite{Droghetti.17}.  The impurity energy is set to $\epsilon_f=-0.13$ (eV) and the temperature is T=$5$ K. (Units are in eV).} 
\end{center}
\end{figure}

\begin{figure}[!t]
\begin{center}
\includegraphics[width=1\columnwidth]{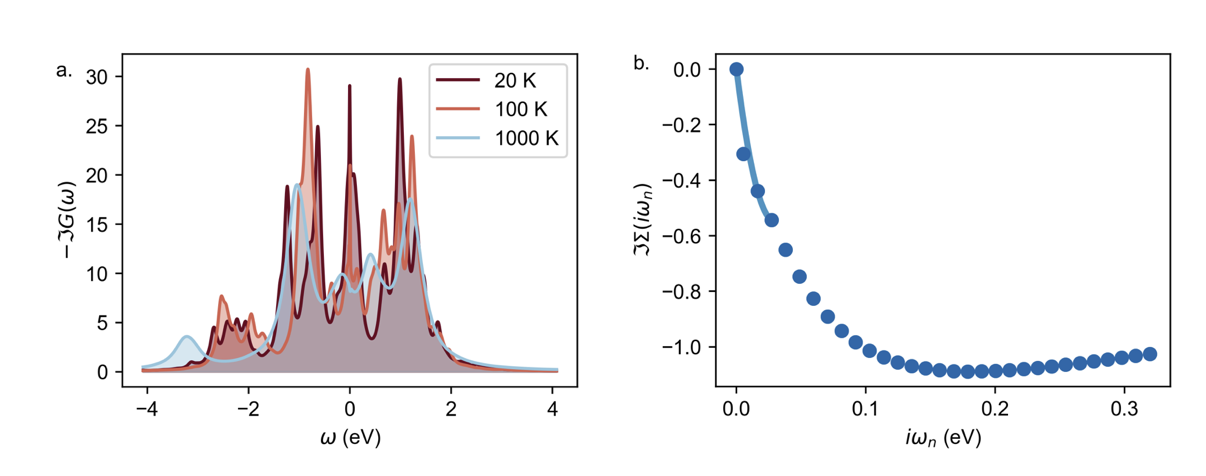}
\caption{\textbf{TOV molecule on Gold surface}. a)Density of states as function of real energies for different temperatures and b) data points of the self-energy at $T=20 K$ (circles) fitted at low frequencies with a quadratic function.} 
\end{center}
\end{figure}


\section{DFT+DMFT for LaNiO$_3$}
DFT+DMFT calculation is performed within Questaal package \cite{Pashov_2020}. We used the local density approximation (LDA) functional with a k-mesh  10x10x10 points. The DMFT subspace is defined by the projection on Ni-d orbitals. The filled $t_{2g}$ manifold is treated at the Hartree Fock level, and the $e_{g}$ is treated using DMFT. In order to solve the Anderson impurity model composed of $e_g$ orbital, we use MLDE with 2 impurity sites and 3 sites in the bath. The comparison with an exact solver is done using CTQMC solver in TRIQS Package \cite{Seth2016274}. 
The interaction  between the 2 $e_g$ orbitals is given by the rotationally invariant Slater Hamiltonian

\begin{equation}
  H_{int} = \frac{1}{2} \sum_{ijlk,\s\s'} U_{ijkl}c^{\dd}_{i\s}    c^{\dd}_{j\s'}    c_{k\s'}    c_{l\s}
\end{equation}

The tensor $U_{ijkl}$ is given in term of Slater integral
\begin{equation}
  U_{ijkl}=\sum^{2}_{k=0}\a_k^{ijlk}F^{2k}
\end{equation}
Where $\a_k^{ijlk}$ are the Racah-Wigner numbers and $F_k$ are the Slater integrals.

In the literature, we often use the parameter corresponding to  the Hubbard interaction $U$ and Hund's coupling $J$ to express $F^{k}$ with the relations

\begin{align}
  U &= F^0\\
  J &= \frac{F^2+F^4}{14}\\
  \frac{F^4}{F^2} &= 10/16
\end{align}

In this letter, we use the parameters $U=7$ and $J=1$ as suggested in \cite{PhysRevB.92.245109}.
\section{MLDE quantum computing algorithm}

In this section, we present a quantum algorithm to run MLDE on noisy intermediate-scale quantum (NISQ) computers. We perform a Jordan-Wigner (JW) transformation \cite{PhysRevLett.63.322} to map the MLDE Hamiltionian (Eq. \ref{eq:Hmlde}) to a qubit representation. After the transformation, the Hamiltonian is in the form of a linear combination of Pauli tensors. With the JW transformation, the required number of qubits is two times the number of sites in the system, the factor two being due to spin.

To compute the Green's function, we use a variational quantum eigensolver (VQE) based approach as presented in \cite{rungger2020dynamical} since it is rather resilient to noise. 
The quality of the quantum computing solution corresponds largely to the ability of the quantum circuit to represent the eigenstates of $H$, and hence obtain accurate energies and amplitudes of the peaks in the Green's function. One, therefore, needs a state preparation ansatz able to accurately represent the ground state as well as the excited states with one electron added or removed \cite{rungger2020dynamical}. For the results presented in the main manuscript, we use a so-called Hamiltonian variational ansatz (HVA) \cite{Wiersema_2020} to represent the eigenstates of $H$. Note that we obtain a similar accuracy also with different types of circuit ansatz, such as a so-called hardware efficient ansatz\cite{HardwareEffAns}. A general quantum state $\left.|\psi(\pmb{\theta})\right>$ is prepared with a circuit ansatz as $\left.|\psi\pmb{\theta}\right>=\hat U(\pmb{\theta}) \left.|0\right>$, where $\left.|0\right>$ is the initial state corresponding to all qubits set to the zero state, and $\hat U(\pmb{\theta})$ is the unitary generated by the state preparation quantum circuit. Here $\pmb{\theta}$ is a vector of parameters, which determine the specific state generated by the unitary. For the HVA $U(\pmb{\theta})$ has the form
\begin{equation}
    \hat U(\pmb{\theta}) = \prod_k^{n_{\text{layers}}} \prod_j e^{-i \theta^k_j \hat P_j} 
\end{equation}
where $P_j$ is a Pauli tensor in $H$ after JW transformation and $\theta^k_j$ a real-valued parameter. The multiplication over $j$ goes over all Pauli terms in the Hamiltonian. The ansatz corresponds to $n_\mathrm{layers}$ repetitions of individual blocks. In a given block, each Pauli tensor term in $H$, $P_j$, is included in the ansatz as imaginary exponentiation with the parameter $\theta^k_j$, which can be readily implemented on a quantum computer \cite{Wiersema_2020}. The accuracy of the ansatz generally improves with increasing $n_\mathrm{layers}$. We verified that for the system considered in the main manuscript, $n_\mathrm{layesr}=2$ gives good accuracy.

To obtain the state preparation circuit parameters for the ground state, we minimize the cost function corresponding to the ground state energy
\begin{equation}
E_\mathrm{GS} = \mathrm{min}_{\pmb{\theta}} E(\pmb{\theta})  = \mathrm{min}_{\pmb{\theta}} \mean{\psi(\pmb{\theta})|H|\psi(\pmb{\theta})}.
\end{equation}
To obtain the state preparation circuit parameters for the excited states there are several possibilities \cite{PhysRevResearch.1.033062,Higgott2019variationalquantum,PhysRevA.99.062304}. Here we use the method proposed  in \cite{Higgott2019variationalquantum}, where a penalty term is added to the Hamiltonian to impose orthogonality of each excited state to the lower energy states. To calculate the $l^\mathrm{th}$ excited state, $\left.|\psi_l\right>=\left.|\psi(\pmb{\theta}_l)\right>$, with energy $E_l$, one therefore minimises
\begin{equation}
E_{l} = \mathrm{min}_{\pmb{\theta}}\left[ \mean{\psi(\pmb{\theta})|H|\psi(\pmb{\theta})} + \sum^{l-1}_{j=0} \a_j\mean{\psi_j|\psi(\pmb{\theta})}^2\right].
\end{equation}
Here the positive-valued $\alpha_j$ are arbitrary parameters, set in a way to optimize the convergence of the algorithm. 
Once all required state preparation parameters $\pmb{\theta}_l$ are computed, the Green's function is obtained using the Lehman representation, and with it, the self-energy \cite{rungger2020dynamical}.


\end{document}